\documentclass[11pt,a4paper]{article}
\pdfoutput=1
\usepackage{jheppub}
\usepackage[T1]{fontenc}
\usepackage{braket}
\usepackage{amsmath}
\usepackage{slashed}
\usepackage{verbatim}
\usepackage[titletoc]{appendix}
\usepackage{graphicx}
\usepackage{sidecap}
\usepackage{amsthm}
\usepackage{bbm}
\usepackage{color}
\usepackage{subcaption}
\usepackage{amsthm}
\usepackage{amssymb}
\newtheorem{thrm}{Theorem}
\newcommand{\Tr}{\mathrm{Tr}}
\newcommand{\Id}{\mathrm{Id}}

\begin {document}

\title{\bf Learning the Alpha-bits of Black Holes}
\author{Patrick Hayden}
\author{and Geoffrey Penington}
\affiliation{\small \em  Stanford Institute for Theoretical Physics, Stanford University, Stanford CA 94305 USA}
\abstract{When the bulk geometry in AdS/CFT contains a black hole, boundary subregions may be sufficient to reconstruct certain bulk operators if and only if the black hole microstate is known, an example of state dependence. Reconstructions exist for any microstate, but no reconstruction works for all microstates. We refine this dichotomy, demonstrating that the same boundary operator can often be used for large subspaces of black hole microstates, corresponding to a constant fraction $\alpha$ of the black hole entropy. In the Schr\"{o}dinger picture, the boundary subregion encodes the $\alpha$-bits (a concept from quantum information) of a bulk region containing the black hole and bounded by extremal surfaces. These results have important consequences for the structure of AdS/CFT and for quantum information. Firstly, they imply that the bulk reconstruction is necessarily only approximate and allow us to place non-perturbative lower bounds on the error when doing so. Second, they provide a simple and tractable limit in which the entanglement wedge is state dependent, but in a highly controlled way. Although the state dependence of operators comes from ordinary quantum error correction, there are clear connections to the Papadodimas-Raju proposal for understanding operators behind black hole horizons. In tensor network toy models of AdS/CFT, we see how state dependence arises from the bulk operator being `pushed' through the black hole itself. Finally, we show that black holes provide the first `explicit' examples of capacity-achieving $\alpha$-bit codes. Unintuitively, Hawking radiation always reveals the $\alpha$-bits of a black hole as soon as possible. In an appendix, we apply a result from the quantum information literature to prove that entanglement wedge reconstruction can be made exact to all orders in $1/N$.}

\emailAdd{phayden@stanford.edu}
\emailAdd{geoffp@stanford.edu}
\maketitle
\flushbottom


\section{Introduction}
Recent work \cite{almheiri2015bulk, harlow2017ryu} has made clear that the semiclassical limit of the AdS/CFT duality is perhaps best understood in the language of quantum error correcting codes.\footnote{This is just one of a series of insights that have been reached about quantum gravity through applying the tools of quantum information \cite{hirata2007ads, van2010building, verlinde2013black,lloyd2014unitarity, yoshida2017efficient}.}  Specifically, the Hilbert space of certain conformal field theories contains subspaces of states (we shall refer to these as the bulk Hilbert spaces or code spaces), which have a dual interpretation as a quantum field theory on a semiclassical gravitational background that is asymptotically anti-de Sitter space. The error correcting codes are the isometries from the bulk subspaces to the larger boundary space.

It is generally understood that the error correction should only become exact in the limit of vanishing Newton's constant $G_N \to 0$ or, equivalently, diverging gauge group rank $N \to \infty$. Nonetheless, out of convenience, many of the toy models and mathematical results \cite{pastawski2015holographic,dong2016reconstruction,harlow2017ryu} that have been developed for understanding the AdS/CFT error-correction paradigm involve finite dimensional Hilbert spaces with exact quantum error correction. In this paper, we show that such a framework is insufficient to capture some of the important aspects of error correction in AdS/CFT, which are only possible for finite-dimensional code spaces if the error correction is merely approximate, rather than exact. 

The magnitude $\varepsilon$ of these uncorrectable errors is very small -- in fact they are non-perturbatively suppressed in the semiclassical limit (their Taylor expansion in $G_N$ or $1/N$ has no non-zero terms). However, we show that the existence of these tiny, seemingly insignificant approximations makes possible key features of the AdS/CFT correspondence that provably could not otherwise exist.

Such features arise when the code space $\mathcal{H}_{\text{code}}$ of `nice' states with smooth bulk geometries that we wish to be able to error correct includes states containing black holes. Since the Bekenstein-Hawking entropy becomes infinite in the classical limit $G_N \to 0$, the maximum dimension of a code space consisting of a large number of black hole microstates diverges, if the horizon area is held fixed in AdS units. As a result, there is no limit in which the error $\varepsilon \to 0$, the code space contains `all' the black hole microstates (or even contains a constant fraction of the black hole entropy), and the dimension of the code space remains bounded. This should immediately make us cautious about believing that results based on the twin assumptions of exact error correction and finite-dimensional Hilbert spaces will continue to be valid in this context.

Suppose we consider a black hole in AdS/CFT, together with a boundary region $A$ that consists of slightly over half of the entire boundary. The size of the entanglement wedge of region $A$ (the region of the bulk that can be reconstructed from boundary region $A$) depends on whether the black hole is in a specific known microstate or if the black hole is an unknown state (which we can model as the thermal ensemble). For the known microstate, the black hole is contained in the entanglement wedge of $A$; for the thermal ensemble, it lies outside. Because the two entanglement wedges are not the same, there exist bulk operators that can be reconstructed in boundary region $A$ if and, more importantly, \emph{only} if the state of the black hole is known. Such operators exist for every black hole microstate, but there exists no single boundary operator that works for all microstates. In other words, the boundary operator is state-dependent.

We emphasize that this definition of state-dependent operator reconstruction should be distinguished from the conjecture, advocated for most prominently by Papadodimas and Raju \cite{papadodimas2013infalling, papadodimas2014state}, that interior operators are necessarily state dependent (although there appear to be close connections between the two, see Section \ref{sec:statedepend}). In the Papadodimas-Raju constructions, even a global reconstruction (that is allowed to act on the entire boundary) is necessarily state dependent. In contrast, in our case, state dependence is only necessary when the reconstruction is restricted to acting on a subregion of the boundary. All our results can therefore be completely understood in the framework of quantum error correction; there is no conflict with the linearity of quantum mechanics.

More specifically, the existence of state-dependent operators for all states (but the absence of \textit{state-independent} operators) corresponds in the Schr\"{o}dinger picture to a weakened version of the usual notion of a quantum error correcting code, called universal subspace quantum error correction. This was studied in detail in \cite{alphabits}. In terms of the language introduced in \cite{alphabits}, we say that region $A$ encodes the \textit{zero-bits} of the bulk region.

\begin{figure}[t]
\includegraphics[width = 0.5\linewidth]{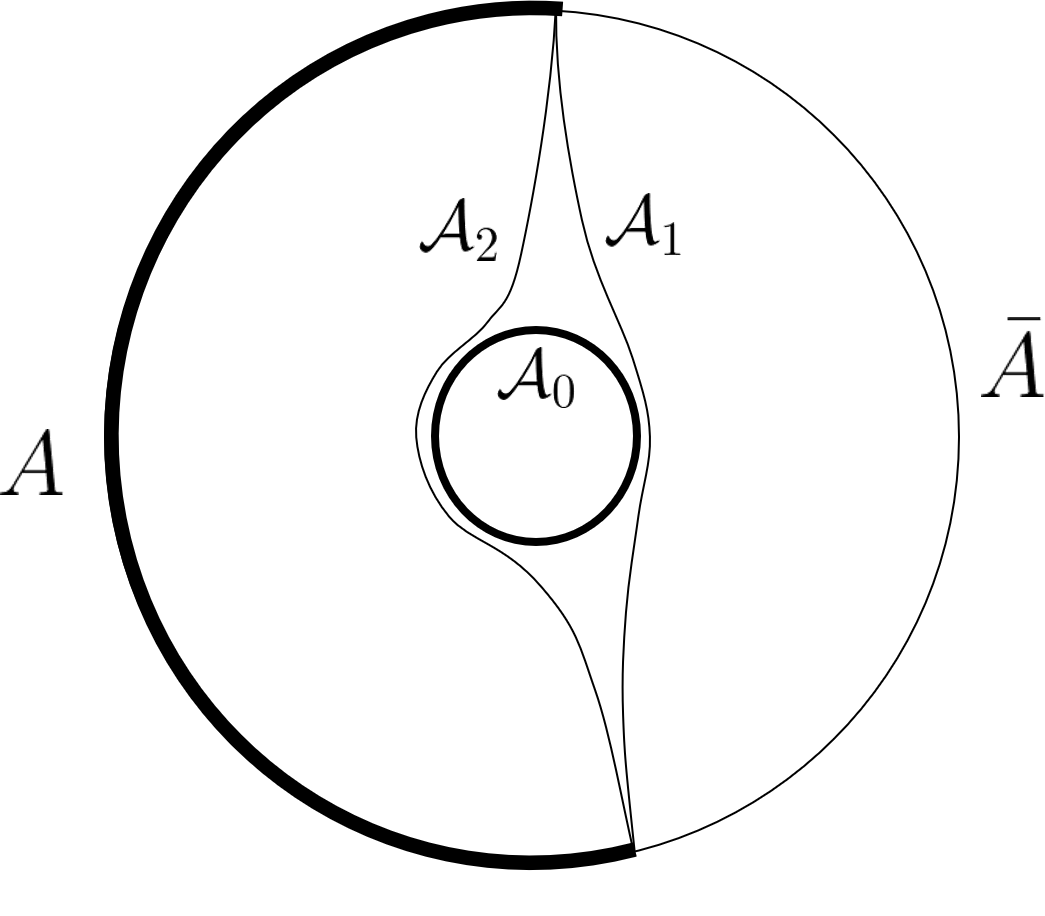}
\centering
\caption{A black hole with horizon area $\mathcal{A}_0$ in AdS-space. The boundary is separated into two regions, $A$ and $\bar{A}$ with shared boundary $\partial A$. There are two important bulk minimal surfaces with boundary $\partial A$. The minimal surface homologous to $\bar A$ has area $\mathcal{A}_1$, while the minimal surface homologous to $A$ has area $\mathcal{A}_2$.}
\label{fig:bhintro}
\end{figure}

As the size of the boundary region $A$ increases, less state dependence is required. A single boundary operator can now reconstruct bulk operators for code subspaces containing many different black hole microstates, although the reconstruction will still necessarily depend on the code subspace chosen.\footnote{We shall continue to use the term ``state dependence'' even when a single reconstruction can simultaneously work for many, but not all, states, because the alternative expression ``code subspace dependence'', while more precise, is something of a mouthful.} As shown in Figure \ref{fig:bhintro}, in general there will be two minimal surfaces with boundary $\partial A$ that might form the RT surface for states in this code space. One will be homologous to $\bar A$ (and hence contain the black hole in the entanglement wedge of $A$), while the other will be homologous to $A$. We say that they have areas $\mathcal{A}_1$ and $\mathcal{A}_2$ respectively; if region $A$ is greater than half the boundary, then $\mathcal{A}_2 > \mathcal{A}_1$.

We show that there exists a single boundary reconstruction in region $A$ of a given bulk operator that works for every state in a subspace, so long as the bulk operator lies within the entanglement wedge of $A$ for every state in the subspace, including states that are entangled with a reference system. Equivalently, the bulk operator must lie inside the entanglement wedge of $A$ for all states \emph{including mixed states} with support only in the subspace. 

If there were no black hole, this subtle distinction of requiring even mixed states to contain the bulk operator in their entanglement wedge would be unimportant. The Ryu-Takayanagi surface (or more precisely the quantum extremal surface \cite{engelhardt2015quantum}), which bounds the entanglement wedge, is defined as the surface that minimises the sum of $\mathcal{A}/4G_N$ (where $\mathcal{A}$ is the area of the surface) and the bulk entropy $S_{\text{bulk}}$. If the bulk entropy
$$ S_{\text{bulk}} = O(1), $$
this will always be the surface with area $\mathcal{A}_1$, at least in the limit $G_N \to 0$. However, if we consider a subspace of black hole microstates of sufficiently large dimension $d$ such that
\begin{align}
\log d + \frac{\mathcal{A}_1}{4 G_N} = \max (S_{\text{bulk}}) +  \frac{\mathcal{A}_1}{4 G_N} >  \frac{\mathcal{A}_2}{4 G_N},
\end{align}
then the RT surface can jump to the surface with area $\mathcal{A}_2$ for states sufficiently entangled with the reference system (or sufficiently mixed). For such states, the entanglement wedge of $A$ will no longer contain either the black hole or the bulk region between the two minimal surfaces but outside the black hole. It is not possible to find a single boundary operator reconstruction for operators in this region that will work for that entire subspace. However, if the dimension $d$ is not sufficiently large for this to happen, then the entanglement wedge will always lie on the original surface. Hence, operators between the minimal surfaces (as well as operators acting on the black hole itself) can be reconstructed from the boundary region $A$.

In other words, any bulk operator lying between the two minimal surfaces can be reconstructed as a single boundary operator that works for \textit{any} subspace of black hole microstates of dimension $d < e^{\alpha S}$ where $S = \mathcal{A}_0 / 4 G_N$ is the Bekenstein-Hawking entropy. The dimensionless parameter 
$$\alpha = \frac{\mathcal{A}_2 - \mathcal{A}_1}{\mathcal{A}_0}$$ is independent of $G_N$, so remains fixed in the semiclassical limit. (In accordance with our previous claim, it should be clear that the parameter $\alpha$ increases as the size of region $A$ increases, until eventually $\alpha = 1$ and a single operator can work for all black hole microstates.)
As above, this is an example of universal subspace quantum error correction. The difference now is that the dimension of the subspace which can be error-corrected is allowed to grow with the dimension of the larger space of all black hole microstates. If we again make use of terminology from \cite{alphabits}, the region $A$ now encodes the \textit{$\alpha$-bits} of the bulk region.

Nontrivial realisations of universal subspace quantum error correction are only possible when the error correction is approximate. In the exact setting, being able to correct all small subspaces automatically implies being able to correct arbitrary subspaces.
Even if the error correction is approximate, the error in correcting subspaces of larger dimension can be bounded in terms of the error for subspaces with smaller dimension. However, the quality of the approximation degrades as the dimension of the subspaces increases.
This makes it possible to find a limit in which the errors tend to zero for small subspaces, but stay order one for large subspaces, so long as in this limit the dimension of the full code space tends to infinity. 
We will see that the classical limit $G_N \to 0$ ($N \to \infty$) in AdS/CFT is an example of precisely this kind of limit. The seemingly insignificant, non-perturbatively small errors make possible order one effects that continue to exist, and in fact become more sharply defined, even in the semiclassical limit.

In Section \ref{sec:abits}, we review the basic construction of $\alpha$-bits and universal subspace error correction. Section \ref{sec:hawking} then shows that the evaporation of black holes into Hawking radiation provides a natural example of a capacity-achieving $\alpha$-bit code. In contrast to our usual intuition, black holes rush to reveal their $\alpha$-bits in the Hawking radiation as quickly as they possibly can; in this sense the Hawking radiation contains as much (rather than as little) information as possible about the state of the black hole. Earlier work by Hayden and Preskill in \cite{hayden2007black} on information retrieval from evaporating black holes can be interpreted as the $\alpha = 0$ case of this more general fact.

In Section \ref{sec:entwedge}, we develop the main result of the paper: the appearance in AdS/CFT of $\alpha$-bit encodings for code spaces containing black holes. We develop the ideas sketched out above in significantly greater depth and precision. Section \ref{sec:btz} includes more specific calculations for the simple case of a uncharged, non-rotating BTZ black hole in AdS$_3$. They show that the region between the minimal surfaces but outside the black hole horizon is always approximately AdS scale, regardless of the size of the black hole -- at least in this simple case, it requires a large central charge CFT with a weakly curved gravity dual to have locality at small scales compared to the size of this `$\alpha$-bit' region. 

Section \ref{sec:tensor} explores how $\alpha$-bit codes can arise in a basic tensor network toy model of AdS/CFT. In this context, the intuition behind the state dependence of the boundary operators can be made very clear; bulk operators have to be pushed through the black hole itself in order to reach the boundary, in a way that manifestly depends on the subspace of black hole microstates being considered. Meanwhile, Section \ref{sec:microspace} provides more detailed, technical justifications that back up our assumptions about the existence of a code space of black hole microstates with the correct entropy.

Section \ref{sec:discuss} consists of an extended discussion on various aspects and implications of the paper. This discussion makes use of many of the results derived in the main sections of the paper, but can be read relatively independently. We show how $\alpha$-bit codes can be used to put lower bounds on the uncorrectable error $\varepsilon$. Specifically, even though it is possible to make the error equal to zero to all orders in perturbation theory, we show that there must sometimes exist errors
$$\varepsilon > e^{-\eta/G_N}$$
for any $\eta > 0$. We argue that $\alpha$-bit codes provide the most controlled setting in which we can understand a state-dependent entanglement wedge, before discussing tantalising connections and similarities between the $\alpha$-bit codes we discuss and the Papadodimas-Raju proposal \cite{papadodimas2013infalling, papadodimas2014state} for the state dependence of operators behind the black hole horizon. We then briefly discuss explicit recovery maps for the $\alpha$-bit codes and, lastly, observe that black holes provide examples of `explicit' (as opposed to randomly generated) capacity-achieving $\alpha$-bit codes for noiseless quantum channels, which until now had not been known.

Finally Appendix \ref{sec:algebra} describes a result from quantum information that has so far not appeared in the quantum gravity literature \cite{beny2009conditions}. It provides a generalisation of the Dong-Harlow-Wall condition \cite{dong2016reconstruction} to approximate reconstruction and is sufficient to prove that entanglement wedge reconstruction can be made exact to all orders in $1/N$. A special case of this result is used in Section \ref{sec:entwedge}, but the theorem and some background are included in full because of the relevance to the wider literature on quantum error correction and AdS/CFT.

\section{Alpha-bits} \label{sec:abits}
We begin with a basic review of the concept of universal subspace error correction and $\alpha$-bits. For more detail see \cite{alphabits}. Because the definitions involved are quite technical, we will begin by illustrating the basic phenomenon we are trying to capture with a relatively simple example.

\begin{figure}[t]
\includegraphics[width = 0.7\linewidth]{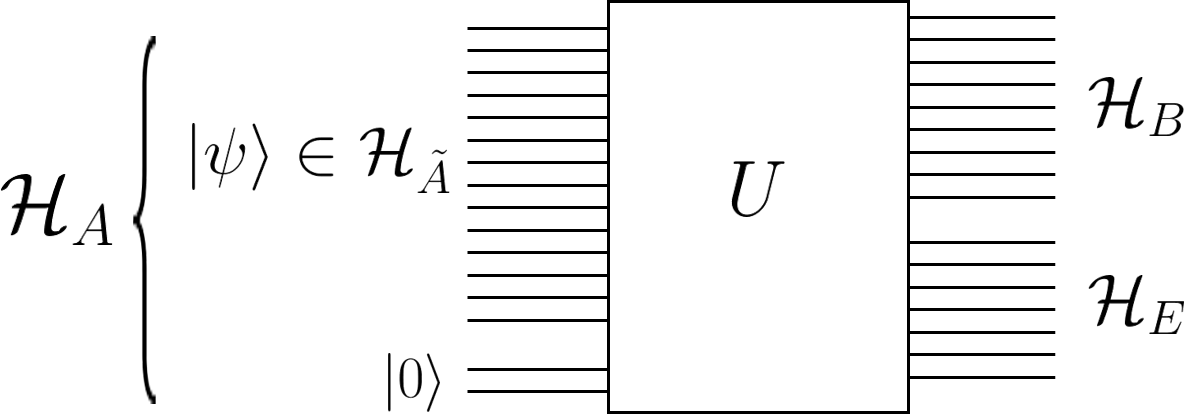}
\centering
\caption{Alice has a quantum state $\ket{\psi} \in \mathcal{H}_{\tilde{A}}$ consisting of $n$ qubits, for some large $n$. She adds a few qubits in a fixed state $\ket{0}$ (embedding $\mathcal{H}_{\tilde A}$ as a subspace of a slightly larger Hilbert space $\mathcal{H}_A$), applies a Haar random unitary $U$ and then sends slightly more than half of the qubits to Bob, thowing the rest away. We say that Alice has sent the zerobits of the state $\ket{\psi}\in \mathcal{H}_{\tilde{A}}$ to Bob.}
\label{fig:zerobits}
\end{figure}

Suppose we apply a Haar-random unitary $U$ to some large number $n$ of qubits then throw away some fraction that is slightly less than half, as shown in Figure \ref{fig:zerobits}. Call the input Hilbert space $\mathcal{H}_A$, the qubits that are kept $\mathcal{H}_B$ and the qubits that are discarded $\mathcal{H}_E$. Now consider the fate of a typical pair of orthogonal pure states on $\mathcal{H}_A$ in the limit of large $n$. Both will get mapped to states almost maximally entangled between $\mathcal{H}_B$ and $\mathcal{H}_E$. Moreover, because $\mathcal{H}_E$ is much smaller than $\mathcal{H}_B$, the reduced states on $\mathcal{H}_E$ will be nearly maximally mixed and therefore effectively indistinguishable. For the same reason, the states on $\mathcal{H}_B$ will have small rank relative to the dimension of $\mathcal{H}_B$, which leads to their being nearly orthogonal.

In fact, due to strong measure concentration effects in high dimension, those properties will hold not just for one pair of orthogonal states on $\mathcal{H}_A$, or two pairs, or even a countable number of pairs. It will hold for \emph{all} pairs of orthogonal states in a subspace $\mathcal{H}_{\tilde A} \subseteq \mathcal{H}_A$ that is almost as large as $\mathcal{H}_A$ in qubit terms: $n - o(n)$ qubits. More generally, the map from $\mathcal{H}_{\tilde A}$ to $\mathcal{H}_B$ approximately preserves the pairwise distinguishability of states as measured by the trace distance despite shrinking the number of qubits by a factor of two~\cite{winter:q-ID-1,hayden2012weak}. Because the dimension of  $\mathcal{H}_B$ is roughly the square root of the dimension of $\mathcal{H}_{\tilde A}$, this seems paradoxical. The resolution is that the map encodes the geometry of the unit sphere of the Hilbert space $\mathcal{H}_{\tilde A}$ into the space of \emph{density matrices} on $\mathcal{H}_B$; the pure state geometry is partially encoded into noise. ``Sending the zero-bits of $\mathcal{H}_{\tilde A}$'' can be roughly defined as approximately preserving the geometry of the unit sphere of pure states. We shall provide a precise technical definition in Section \ref{sec:tech}.

Returning to the example, while the full space $\mathcal{H}_{\tilde A}$ has been transmitted in some sense, it is clearly not possible to perform approximate quantum error correction and completely reverse the effect of the channel; doing so would lead to the quantum capacity of a qubit being greater than one, which by recursion would make it infinite. Geometry preservation does have an operational consequence, however. If we restrict the states to \emph{any} two-dimensional subspace of $\mathcal{H}_{\tilde A}$, there is a decoding operation that will perform quantum error correction. The only catch is that the decoding operation will in general depend on the two-dimensional subspace that we wish to decode. Note, however, that the encoding $\mathcal{H}_{\tilde A} \hookrightarrow \mathcal{H}_A$ and the channel do not. If we think of Alice sending Bob a state, then Bob has to know which two-dimensional subspace the state is in, but Alice does not. 

What if Bob wishes to be able to decode larger subspaces of $\mathcal{H}_{\tilde A}$? What fraction of the qubits need to be kept now? As one would expect, to decode the entire space $\mathcal{H}_{\tilde A}$ they have to keep almost all of them. On the other hand, so long as they keep a fraction greater than $\frac{1+\alpha}{2}$ of the qubits, Bob can decode any subspace of up to $\alpha n$ qubits (or in other words any subspace of dimension at most $2^{\alpha n}$).\footnote{Technically, the construction in \cite{alphabits} requires the use of shared randomness to achieve this rate but this can be eliminated by block coding.} We call the task of decoding any such subspace \emph{universal approximate subspace error correction} and say that $\mathcal{H}_B$ encodes the $\alpha$-bits of $\mathcal{H}_{\tilde A}$.\footnote{Note that the \emph{number} of $\alpha$-bits sent is determined by the dimension of $\mathcal{H}_{\tilde A}$ rather than the dimension of the subspaces Bob wishes to decode. This is because the whole space $\mathcal{H}_{\tilde A}$ is available to him; he just needs to make a choice about which subspace he is interested in. Furthermore, in the case of zero-bits, the subspace dimension is always $d=2$, even though the amount of information sent clearly depends on the size of $\mathcal{H}_{\tilde A}$, and so this is the only sensible definition.} A zero-bit is then simply the special case of an $\alpha$-bit with $\alpha = 0$.\footnote{Since decoding a one-dimensional state is trivial, we need the subspaces to have dimension $2^{\alpha n} + 1$ to correctly recover the definition of zerobits when $\alpha = 0$. This change has negligible effect on the definition of $\alpha$-bits for $\alpha > 0$; at most it can slightly increase the size of the error in recovering the state.}

\subsection{Technical definitions} \label{sec:tech}
We now turn to a more formal definition. Readers satisfied by the level of rigour given above should feel free to proceed to the next section. An exact quantum error correcting code consists at its simplest of an encoding and transmission channel $\mathcal{N}: S(\mathcal{H}_A) \to S(\mathcal{H}_B)$ (where $S(\mathcal{H}_A)$ is the space of density matrices on $\mathcal{H}_A$) together with a decoding or recovery channel $\mathcal{D}:S(\mathcal{H}_B) \to S(\mathcal{H}_A)$ such that $$\mathcal{D} \circ \mathcal{N} = \Id_A.$$ In other words, given any state $\rho \in S(\mathcal{H}_A)$, it is possible to exactly recover the state $\rho$ from the state $\mathcal{N}(\rho)$.

Suppose, as above, that we allow the receiver Bob to have some additional information about the state $\rho$; he again knows that the state lies within some particular subspace of $A$. Obviously this can make the task of finding a recovery channel $\tilde{\mathcal{D}}$ considerably easier. Indeed, the normal approach used to make error correcting codes out of a noisy transmission channel is to first apply an encoding channel that consists of an isometry from a (smaller) code space into the (larger) input space of the transmission channel, in such a way that the code space is possible to decode, even though the large input space is not. 

However, in this case we don't just want Bob to be able to decode some particular subspace that has especially nice properties. We want him to be able to decode the state provided he knows \textit{any} sufficiently small subspace that it is contained in. In the framework of exact quantum error correction, this would mean the existence of an exact decoding channel $\mathcal{D}_S$ for any sufficiently small subspace $\mathcal{H}_S$.\footnote{We emphasize again that the encoding is not allowed to depend on $\mathcal{H}_S$. Otherwise the encoding channel can simply map $\mathcal{H}_S$ to any fixed subspace, and the task becomes identical to ordinary error correction for the smaller space $\mathcal{H}_S$.} However, it turns out that, even if we only require any two-dimensional subspace to be error-correctable, the existence of a decoding channel for every subspace implies that the complete space $\mathcal{H}_A$ can also be error-corrected. In other words there is no advantage to being able to use a different decoding channel for each subspace, if you still have to be able to decode any possible subspace exactly.

How can this be reconciled with our analysis above? The answer, of course, lies in our assumption that the error correction had to be exact. If we instead only require approximate error correction to be possible, the situation becomes completely different. In this case we only require that the decoding channel get back something very close to (rather than exactly) the original state. In other words we only require that, $$\left \lVert \mathcal{D}_S \circ \mathcal{N} - \Id \right \rVert_\diamond \leq \varepsilon$$ for some $\varepsilon \ll 1$.\footnote{We have bounded the error in terms of the diamond norm (see Appendix \ref{sec:algebra}) here, but we could have equally used the operator norm since these bound each other in a dimension-independent way in the neighbourhood of the identity \cite{kretschmann2004tema}.}

The Stinespring dilation theorem says that for any channel $\mathcal{N}: S(\mathcal{H}_A) \to S(\mathcal{H}_B)$ there exists an isometry $V: \mathcal{H}_A \to \mathcal{H}_B \otimes \mathcal{H}_E$ that is unique up to isomorphisms of $\mathcal{H}_E$ such that for all states $\rho \in S(\mathcal{H}_A)$, $$\mathcal{N}(\rho) = \Tr_E V \rho V^\dagger.$$ We can then define the complementary channel $\mathcal{N}^c: S(\mathcal{H}_A) \to S(\mathcal{H}_E)$ by $$\mathcal{N}^c(\rho) = \Tr_B V \rho V^\dagger.$$

The subspace decoupling duality, proved in \cite{alphabits}, states that, if there exists a decoding channel $\mathcal{D}_S$, for any subspace of dimension less than or equal to $k$, with error at most $\varepsilon$ as above, then there exists a state $\sigma \in S(\mathcal{H}_E)$ such that for \emph{all} states $\rho \in S(\mathcal{H}_R\otimes \mathcal{H}_A)$ 
\begin{align} \label{eq:forgetful}
\left\lVert (\mathcal{N}^c \otimes \Id_R) \rho_{AR} - \sigma_E \otimes \rho_{R} \right\rVert_1 \leq \delta
\end{align} 
where $\mathcal{H}_R$ is a reference system whose dimension is also equal to $k$ and $\delta \leq 8 \sqrt{\varepsilon}$.\footnote{The subspace decoupling duality can be derived almost immediately by applying Kretschmann \emph{et al.}'s information-disturbance theorem \cite{kretschmann2006information} to arbitrary subspaces of dimension $k$.} In other words, the environment $\mathcal{H}_E$ encodes almost no information about the state $\rho$. We say that the complementary channel $\mathcal{N}^c$ is approximately $k$-forgetful. Conversely if the complementary channel $\mathcal{N}^c$ is approximately $k$-forgetful with uncertainty $\delta$, then there exists a decoding channel for any subspace of dimension at most $k$ with error $\varepsilon \leq 2 \sqrt{2\delta}$.

How important is the inclusion of a reference system with dimension $k$ in \eqref{eq:forgetful}? By writing out Schmidt decompositions and using triangle inequality (see for example Lemma 23 of \cite{hayden2012weak}), one can easily show that, for any $k$,
\begin{align} \label{eq:linear}
\max_{\rho \in S(\mathcal{H}_R\otimes \mathcal{H}_A)} \left\lVert (\mathcal{N}^c \otimes \Id_R) \rho_{AR} - \sigma_E \otimes \rho_{R} \right\rVert_1 \leq k \max_{\rho \in S(\mathcal{H}_A)} \left\lVert \mathcal{N}^c (\rho) - \sigma_E \right\rVert_1.
\end{align}
In other words, $\delta$ grows at most linearly with the reference dimension $k$. Hence, as claimed above, exact universal subspace error correction is indeed equivalent to exact quantum error correction; if $\varepsilon = 0$ for $k=2$, then we also have $\delta = 0$ for $k =2$, and hence $\delta = \varepsilon = 0$ for all $k$. 

However there are examples of quantum channels that saturate the bound \eqref{eq:linear} \cite{hayden2012weak}. This means that, when the dimension of the complete space $\mathcal{H}_A$ is very large, the forgetfulness $\delta$ and reconstruction error $\varepsilon$ can end up being much larger for large subspace dimensions $k$ than for small subspace dimensions. In other words, it may be possible to reconstruct any sufficiently small subspace with very high fidelity, while still being completely impossible to reconstruct the entire space.

To make precise statements without reference to epsilons and deltas, it is generally necessary to consider a limit where the dimension $d$ of the code space tends to infinity, for example the classical limit of the space of black hole microstates. In general, the question of whether the error $\varepsilon \to 0$ will then depend on how the subspace dimension $k$ scales with the dimension $d$ of the complete space.

To take a trivial example, universal subspace error correction for subspaces of dimension at most $$k = \frac{d}{t}$$ for some fixed $t$ is equivalent to conventional error correction (i.e. either both have errors that tend to zero in some limit or neither does). However, if the subspace dimension grows sublinearly with $d$, universal subspace error correction is inequivalent to ordinary error correction. The most natural possiblility to consider is that the dimension of the subspaces grows proportionally to $d^\alpha$ for some $0 \leq \alpha \leq 1$. If universal subspace error correction is possible with vanishing error for such a dimension we say that Bob has the $\alpha$-bits of the state sent by Alice. 

The case $\alpha = 0$, which we call \textit{zero-bits}, corresponds to the ability to do universal subspace error correction for any constant dimension $k$ that is independent of $d$. Just as we saw above when $d/k$ was held fixed, the exact value of $k$ does not matter (we generally take $k=2$ for convenience) since, if $\varepsilon \to 0$ for any $k \geq 2$, it will also tend to zero for all fixed $k$, even though, for finite errors, the size of $k$ will affect the size of the error $\varepsilon$. Formally, we define the $\alpha$-bit decoding error based on subspaces of dimension $k = 2^{\alpha n} + 1$, since this formula gives $k=2$ for zero-bits ($k=1$ would be trivial) and scales as $d^\alpha$ for $\alpha > 0$. However, as we have discussed, the exact dimension of the subspace is essentially unimportant; we only care about how this dimension scales with $n$.

\section{Alpha-bits from the Hawking radiation} \label{sec:hawking}
In this section we argue based on simple qubit toy models that an evaporating black hole reveals its $\alpha$-bits through its Hawking radiation as quickly as possible, saturating the $\alpha$-bit capacity of a noiseless quantum channel. This is in sharp contrast with the usual notion that the Hawking radiation tries to hide information about the black hole state for as long as possible, but we shall show that the two ideas are not merely reconcilable but in fact equivalent. We generalise the arguments made by Hayden and Preskill in \cite{hayden2007black}, which can be interpreted as describing the special case where $\alpha = 0$. This section is relatively self-contained and is not necessary to understand the main claims of the paper, which are developed in Section \ref{sec:entwedge}; however, it is both of interest in its own right and features strong similarities to the way the $\alpha$-bits of black holes are encoded in AdS/CFT -- suggesting that the lessons from AdS/CFT may well be important in a significantly broader context.

It is often incorrectly implied that the Hawking radiation contains no information until the Page time, after which it begins to reveal the qubits of the black hole one by one. In fact there is good reason to think that (at least in simple toy models) not a single qubit of the black hole will be revealed, to an observer knowing nothing about the original black hole state, until the black hole has almost entirely evaporated.\footnote{More precisely, there will no tensor product factorisation of the black hole $\mathcal{H}_{BH} \cong \mathcal{H}_{\text{qubit}} \otimes \mathcal{H}_{\text{rest}} $ such that the reduced state on $\mathcal{H}_{\text{qubit}}$ can be determined from the Hawking radiation before the black hole has almost evaporated, even if the dimension of $\mathcal{H}_{\text{qubit}}$ is only two.} Instead, after the Page time, the $\alpha$-bits of the entire black hole will be revealed for increasing values of $\alpha$, until eventually all the qubits are revealed, essentially simultaneously, at the very end of the evaporation process. No particular subsystem is revealed before any other subsystem; however, increasingly large subspaces of the entire system become decodable.

Consider a large semiclassical black hole $A$ in a pure microstate that is \textit{already known} by some observer Bob. Alice wants to hide her diary $D$, a small quantum state, from Bob by dropping it into this black hole. After she has done so, Bob knows that the black hole is in some particular small-dimensional subspace of the large Hilbert space of black hole microstates -- specifically the subspace of states that could have been created by the diary falling in. In the semiclassical limit, the dimension of the space of black hole microstates tends to infinity, while the dimension of the small subspace remains fixed. 

The black hole is then allowed to evaporate into Hawking radiation. We assume Bob has a perfect understanding of the microscopic dynamics of the black hole and the ability to collect all the Hawking radiation that is emitted by it, as well as infinite computational power. However, even with these awesome powers, he has no ability to measure the internal black hole degrees of freedom themselves. How long does Bob have to wait in order to determine the original state of the diary with a high degree of confidence?

This question was studied in detail in \cite{hayden2007black}. Since the dynamics of the black hole interior are expected to be highly chaotic, Bob cannot hope for the small subspace of possible black hole states that could have been created by the diary to be especially easy to decode from the Hawking radiation. The problem is essentially equivalent to the question of whether Bob is able to decode \emph{any} arbitrary small subspace of black hole states. In our language, Bob needs to have access to the \textit{zero-bits} of the black hole.

To conclusively answer the question of how information is encoded in the Hawking radiation (we shall assume throughout this paper that the evaporation is unitary), we would have to understand the exact details of the dynamics of the evaporation of black holes; these details are, of course, as yet unknown. However, considerable progress can be made using some fairly basic assumptions and arguments.

One toy model of the evaporation of a black hole is to add a few ancilla qubits (in order to make the process slightly thermodynamically irreversible) and then to apply a random unitary. However, this is exactly the model that we claimed in Section \ref{sec:abits} had its zero-bits encoded in slight more than half of the output qubits. This suggests that the zero-bits of the black hole should be encoded in any fraction of the Hawking radiation greater than one half. Indeed this is essentially the model and conclusion reached in \cite{hayden2007black}. 

A slightly more sophisticated model of a black hole would be to use an element of a unitary 2-design rather than a fully Haar random unitary. (All the elements of a 2-design can be chosen to be much less computationally complex than a generic Haar random unitary and hence could reasonably be applied within approximately the scrambling time.) Conveniently, so long as we model the black hole as applying a randomly sampled element of the 2-design with the choice of element known by Bob (and not simply as applying a generic element of the 2-design), we obtain exactly the model that was shown in \cite{alphabits} to saturate the $\alpha$-bit capacity for general values of $\alpha$.

Despite their popularity in the literature as toy models of black holes, such random unitary models all suffer from a significant flaw as models of real-life black hole evaporation. Specifically, rather than being only slightly thermodynamically irreversible, black hole evaporation is in general \emph{highly} thermodynamically irreversible, with numerical calculations suggesting that the thermodynamic entropy increases by a factor of approximately $1.48$ over the course of the evaporation process \cite{page2013time}. This has a number of important qualitative effects: for example, it means that the Page time, when the entropy of the radiation equals that of the black hole, occurs significantly before the black hole has lost half its entropy. 

However we can in principle prevent this $O(S)$ thermodynamic entropy increase. For example, we can extract only a small amound of energy and entropy from the Hawking radiation, slightly reducing its temperature, and reflect the rest back into the black hole. Notably, this can be achieved very easily and naturally in AdS/CFT by simply add a weak local coupling to the boundary theory. Alternatively, all but the highest energy Hawking modes (with frequency well above the Hawking temperature) may be reflected back into the black hole by a potential barrier; this happens, for example, in near-extremal Reissner-Nordstr\"{o}m black holes, for example. In the interests of simplicity, we shall therefore assume throughout this section that the black hole evaporates by some close-to-thermodynamically-reversible process (unless stated otherwise).

An alternative argument to the simplified toy models discussed above, but which reaches the same conclusion goes as follows. Rather than make any assumptions about the dynamics of the black hole itself, we can simply assume that the semiclassical result of thermal Hawking radiation is correct -- whenever this assumption is consistent with unitarity. Assuming that our evaporation process is close to being thermodynamically reversible, this implies that the semiclassical calculation should be accurate (and the Hawking radiation should look thermal) so long as we look at less than half of the Hawking radiation. However, the subspace decoupling duality discussed in Section \ref{sec:abits} means that this is equivalent to the zero-bits of the black hole being encoded in any fraction of the Hawking radiation \emph{greater} than half.

A natural generalisation of the problem of reconstructing a small diary thrown into a known black hole is to replace the diary by a second smaller black hole $D$. Now the dimension of the diary Hilbert space is no longer small and fixed; instead it is exponential in $1/G_N$. Unlike the black hole that it is thrown into, the state of the black hole $D$ is unknown to Bob. Let the horizon area of the black hole $D$ be $\alpha \mathcal{A}$ where $\mathcal{A}$ is the horizon area of the final \textit{combined} black hole. 

The subspace of possible black hole states that can be created upon throwing in the diary is no longer small, but instead grows in the semiclassical limit as $e^{\alpha S}$ where $S = \mathcal{A}/4G_N$. To determine the original state of the diary, Bob now needs access to the $\alpha$-bits of the larger combined black hole.

Using the slightly more sophisticated version of the random unitary model, as well as the $\alpha$-bit capacity results from \cite{alphabits}, we see that to recover the $\alpha$-bits of the black hole, Bob needs to obtain at least an $$\frac{\alpha + 1}{2}$$ fraction of the Hawking radiation \cite{alphabits}.

What about if we again try to argue from the principle that the Hawking radiation should look thermal whenever this is consistent with information preservation? To make use of the subspace decoupling duality, we now need to allow the black hole states to be entangled with a reference system $\mathcal{H}_R$ of dimension $e^{\alpha S}$. We want to know whether the reference system $\mathcal{H}_R$, together with the part of the Hawking radiation $\mathcal{H}_E$ which is thrown away, contains any information about the state of the black hole. We assume that the reduced density matrix of the state on $\mathcal{H}_E \otimes \mathcal{H}_R$ will look like the thermal ensemble on $\mathcal{H}_E$ tensored with the reduced density matrix of the original state on $\mathcal{H}_R$, so long as such a state can be purified by the remaining fraction $p$ of the Hawking radiation $\mathcal{H}_B$ which is collected by Bob. This is possible if and only if the dimension of $\mathcal{H}_B$ is larger than the dimension of $\mathcal{H}_E \otimes \mathcal{H}_R$. In other words if $$\alpha + (1-p) \leq p.$$ Hence a fraction $(1-p)$ of the Hawking radiation will be $e^{\alpha S}$-forgetful so long as $\alpha \leq 2p - 1$.

However, by the subspace decoupling duality, if some fraction $1- p$ of the Hawking radiation is $2^{\alpha S}$-forgetful, the remaining fraction $p$ of the Hawking radiation must encode the $\alpha$-bits of the black hole. Rearranging, we again find that the $\alpha$-bits of the black hole are encoded in any fraction $p > (1 + \alpha)/2$ of the Hawking radiation. As for the zero-bit case, the assumption of thermality, whenever it is consistent with unitarity, gives an answer that agrees with the random unitary model.

The black hole evaporation therefore compresses the $\alpha$-bits of the \emph{entire} black hole (consisting of, say, $n$ qubits) into only $(1+\alpha)n/2$ physical qubits of Hawking radiation. The Hawking radiation forms an $\alpha$-bit code for the entire black hole that encodes $$\frac{2}{1+\alpha} \,\,\text{logical}\,\alpha\text{-bits per physical qubit.}$$ This is exactly the $\alpha$-bit capacity of the noiseless qubit channel; black holes give up their $\alpha$-bits as fast as they possibly can. On the one hand, this is unsurprising since random unitary channels were exactly the strategy used in \cite{alphabits} to originally achieve the $\alpha$-bit capacity. It is nonetheless in sharp contrast to usual idea of Hawking radiation containing as little information as possible. 

However, as we have seen, these two phenomena are not merely reconciliable; they are actually equivalent. The subspace decoupling duality means that if the black hole releases as little information as possible in less than half of its Hawking radiation, then it necessarily also releases as much information as possible in more than half of the Hawking radiation (at least in the specific sense of encoding the $\alpha$-bits for as large a value of $\alpha$ as possible). 

Just like a small diary, if a black hole diary (as before with horizon area $\alpha \mathcal{A}$) is thrown into a black hole that has already partially evaporated, the information within it will be revealed more quickly. (Note that $\mathcal{A} = 4 G_N\,S$ is now the horizon area of the combined black hole after it has both been allowed to partially evaporate and then had the black hole diary thrown in.) We can see this for the random unitary model by making use of results used to prove the achievability of the entanglement-assisted $\alpha$-bit capacity in \cite{alphabits}. These show that if, when the black hole diary is thrown in, the black hole is already approximately maximally entangled with Hawking radiation of entropy $\beta  S$, Bob will be able to determine the state of the black hole diary so long as the fraction $p$ of the remaining Hawking radiation that he obtains satisfies
\begin{align}\label{eq:inequal}
\beta S + p S \geq (1-p)S + \alpha S
\end{align}
This inequality can also be derived from the thermality (whenever consistent with unitarity) assumption. The left hand side is the combined entropy of the original Hawking radiation $\beta S$ together with the newly emitted Hawking radiation $p S$ (i.e. the systems that Bob has access to). The right hand side consists of the entropy of the Hawking radiation that is thrown away $(1-p) S$ plus the maximum entropy $\alpha S$ of the reference system $\mathcal{H}_R$ that we need to consider according to (\ref{eq:forgetful}) (i.e. the systems that Bob does not have access to). For these systems to look thermal while being purified by those Bob has access to requires (\ref{eq:inequal}). 

Bob will therefore recover the state of the diary once he has access to at least a 
\begin{align} \label{eq:entangledcapacity}
p = \frac{1+ \alpha - \beta}{2}
\end{align}
fraction of the remaining Hawking radiation. In the special case $\beta = 1$ and $\alpha = 0$, the fraction required tends to zero. This case was studied in detail in \cite{hayden2007black}; it is probable that Bob only needs to wait for at least the scrambling time ($O(\log S)$) before he can recover the state of the diary.

It may seem from \eqref{eq:entangledcapacity} that entangled black holes exceed the entanglement-assisted $\alpha$-bit capacity of $2/(1+\alpha)$ for a noiseless channel that was derived in \cite{alphabits}. However, a precise comparison, done in Section \ref{sec:abitsup}, shows that, as with the unentangled case, it merely saturates the capacity.

At the start of this section, we claimed that, unlike for the $\alpha$-bits of the black holes, we should not expect that even a single qubit of the black hole Hilbert space, no matter what basis we work in, is revealed until almost the end of the Hawking evaporation process. If a qubit was revealed before this point, then there would necessarily exist a pair of states in the black hole Hilbert space that are both maximally entangled with a reference Hilbert space of only one fewer qubit than the black hole Hilbert space, but for which the Hawking radiation produces (almost) orthogonal states after only some fraction $p < 1$ of the Hawking radiation has been produced. Yet for such states, the entropy $(S - \log 2)$ of the reference system, plus the remaining entropy $(1-p)S$ of the black hole is far larger than the entropy $p S$ of the Hawking radiation. It follows that the Hawking radiation should look thermal for randomly chosen states of this form, and hence (again using the fact that our black hole evaporation is slightly thermodynamically non-reversible) measure concentration will be sufficient to ensure that all such pairs of states should have almost indistinguishable Hawking radiation, contradicting our original assumption.

Finally, it is worth commenting briefly on which of our conclusions are likely to change if a black hole is allowed to evaporate \emph{irreversibly}, and which should continue to be valid. In particular, we would expect that the black hole evaporation will no longer saturate the noiseless $\alpha$-bit capacity. For example, the Page time, when the zero-bits of the black hole should be revealed, will now occur when the entropy of the Hawking radiation is significantly more than half of the initial black hole entropy (for realistic black holes the figure is approximately $60\%$ \cite{page2013time}).
\begin{figure}[t]
\includegraphics[width = 0.9\linewidth]{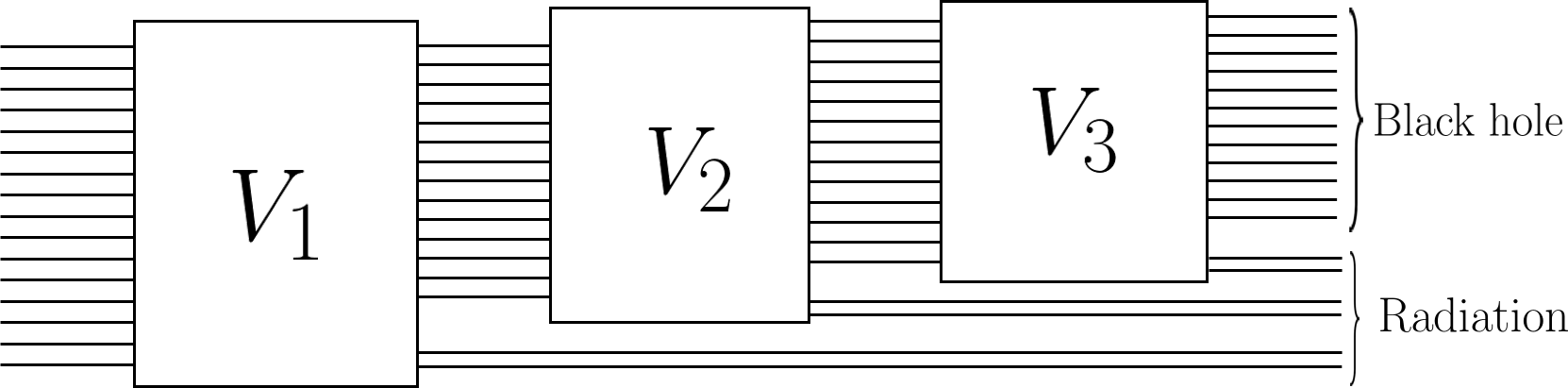}
\centering
\caption{A simple two model of an evaporating black hole that incorporates thermodynamic irreversibility. At each time step, two qubits are released from the black hole as Hawking radiation and then a random isometry is applied to the black hole that adds a single qubit. This model is an example of a random tensor network and hence obeys a version of the Ryu-Takayanagi formula.}
\label{fig:irreversible}
\end{figure}

However, many of the results above should still apply with small adaptations. In particular, it should continue to be the case that the $\alpha$-bits of the black hole are revealed when
\begin{align} \label{eq:irreversibleabitcondition}
\alpha S_0 < S_\text{rad} - S_\text{BH}
\end{align}
where $S_0$ was the initial Bekenstein-Hawking entropy of the black hole, $S_\text{BH}$ is the Bekenstein-Hawking entropy of the partially evaporated black hole and $S_\text{rad}$ is the thermodynamic entropy of the Hawking radiation. 

To see this, note that the natural generalisation of the random unitary models used above to thermodynamically irreversible evaporation is a sequence of nested random isometries. At each step a few qubits are released as Hawking radiation and then a smaller number of qubits are added back to the black hole using a random isometry. Hence, over time, the number of qubits describing the black hole (the Bekenstein-Hawking entropy) decreases, but the total number of qubits describing the black hole together with the Hawking radiation (the total thermodynamic entropy) increases. This is shown in Figure \ref{fig:irreversible} and is an example of a random tensor network; such networks obey a version of the Ryu-Takayanagi formula \cite{hayden2016holographic}, where the entropy of a subsystem is proportional to the number of legs in the network that need to separate the subsystem  from its complement. 

An initial state that is maximally entangled with a reference system with $\log d_R = \alpha S_\text{BH}$ will therefore have maximal entropy (and hence be maximally mixed) on the black hole Hilbert space plus reference system, so long as \eqref{eq:irreversibleabitcondition} holds. By the subspace decoupling duality, the $\alpha$-bits of the initial black hole can therefore be reconstructed from the Hawking radiation at the same point in time. 

The other results in this section can be similarly extended to irreversible evaporation. For example, all the qubits of the initial black hole state are still revealed simultaneously. However this revelation will now happen somewhat before the black hole has completely evaporated, when
\begin{align}
S_\text{rad} - S_\text{BH} > S_0;
\end{align}
the last part of the Hawking radiation can be completely error corrected. 

In general, the inherent inefficiency of an irreversible process prevents the  black hole evaporation from saturating noiseless $\alpha$-bit capacities. However, the random nature of black hole dynamics means that the $\alpha$-bits of the black hole are still revealed `as soon as possible', subject to these inefficiencies.

\section{Alpha-bits in the entanglement wedge} \label{sec:entwedge}

We now turn to developing the main claim of this paper -- that there exist bulk regions, containing but not solely consisting of a black hole, for which the $\alpha$-bits, but only the $\alpha$-bits, are encoded in a certain boundary region. These bulk regions are bounded on both sides by extremal surfaces with areas $\mathcal{A}_1$ and $\mathcal{A}_2$ and they satisfy $\alpha = (\mathcal{A}_2 - \mathcal{A}_1) / \mathcal{A}_0$ where $\mathcal{A}_0$ is the horizon area of the black hole. 

We first introduce the concept of entanglement wedge reconstruction. We then establish a correct version of the entanglement wedge reconstruction conjecture in Section \ref{sec:approxent}, including an underappreciated subtlety that proves qualitatively important for code spaces whose dimension grows quickly in the limit $G_N \to 0$. Finally, in Section \ref{sec:BH} we apply our results to geometries containing a single black hole in AdS space, establishing the results mentioned above. 

The Ryu-Takayanagi formula \cite{ryu2006holographic, faulkner2013quantum} states that, to leading order in $G_N$, the entanglement entropy of a boundary region $A$ is equal to 
\begin{align} \label{eq:rtformula}
S_{RT} = \frac{\mathcal{A}}{4G_N} + S_{\text{bulk}},
\end{align}
where $\mathcal{A}$ is the area of the bulk minimal surface anchored to the boundary of $A$ and $S_{\text{bulk}}$ is the bulk entropy of the bulk region bounded by the boundary region $A$ and the minimal surface. However, as conjectured in \cite{engelhardt2015quantum} and proved in \cite{dong2018entropy}, the correct definition of the `minimal surface' that one must consider is not simply the surface of minimal area (the classical extremal surface),\footnote{We assume throughout this paper that we are in the Einstein gravity limit ($\lambda \to \infty$), where higher curvature corrections can be ignored. Moreover, we only consider static spacetime geometries and hence there is no need to use HRT surfaces \cite{hubeny2007covariant} which generalises RT surfaces (which require a time-reflection symmetry) to general spacetimes. There will no doubt exist $\alpha$-bit codes in time-dependent spacetimes as well, but we do not consider them here.} but rather the quantum extremal surface, which is the surface anchored on the boundary of $A$ that minimises the total size of \eqref{eq:rtformula}.\footnote{To avoid ambiguities about whether homology constraints are being applied the term RT surface will always refer in this paper to the quantum extremal surface.} If the size of the code space is held fixed as $G_N \to 0$, then the second term does not contribute to leading order in $G_N$ and can be safely ignored at this order. However, if the dimension $d_{\text{code}}$ of the code space of allowed bulk states exponentially large in $1/G_N$, the two terms can compete even in the semiclassical limit. This is exactly what happens when one considers thermal or two-sided black hole states; it is one way to understand the source of the `homology constraint' in the original Ryu-Takayanagi formula \cite{harlow2017ryu}.

The conjecture of entanglement wedge reconstruction was developed in \cite{czech2012gravity, headrick2014causality, wall2014maximin} and then established with increasing rigour in~\cite{jafferis2016relative,dong2016reconstruction,cotler2017entanglement}. It states that, if we take a code subspace of states with a fixed geometry, any region $A$ of the boundary acts as a quantum error correcting code for the region of the bulk within its \textit{entanglement wedge}, the bulk region enclosed by the boundary region $A$ and the RT surface (specifically the quantum extremal surface) associated to region $A$. However, as we shall see, the form of the quantum error correction involved is somewhat more general than the definitions that we have given so far. 

\begin{figure} [t]
\centering
\begin{subfigure}{.5\textwidth}
  \centering
 \includegraphics[width = 0.8\linewidth]{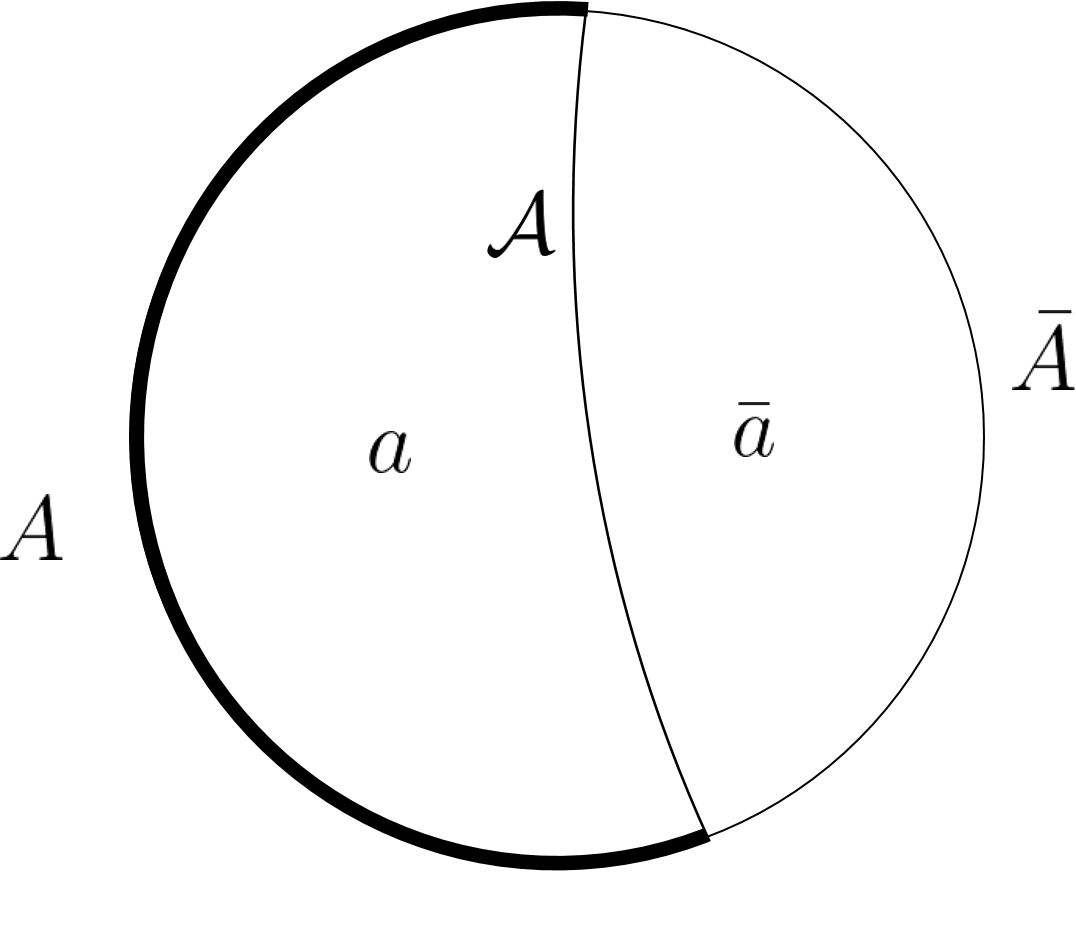}
 \caption{Empty AdS}
\end{subfigure}%
\begin{subfigure}{.5\textwidth}
 \includegraphics[width = 0.8\linewidth]{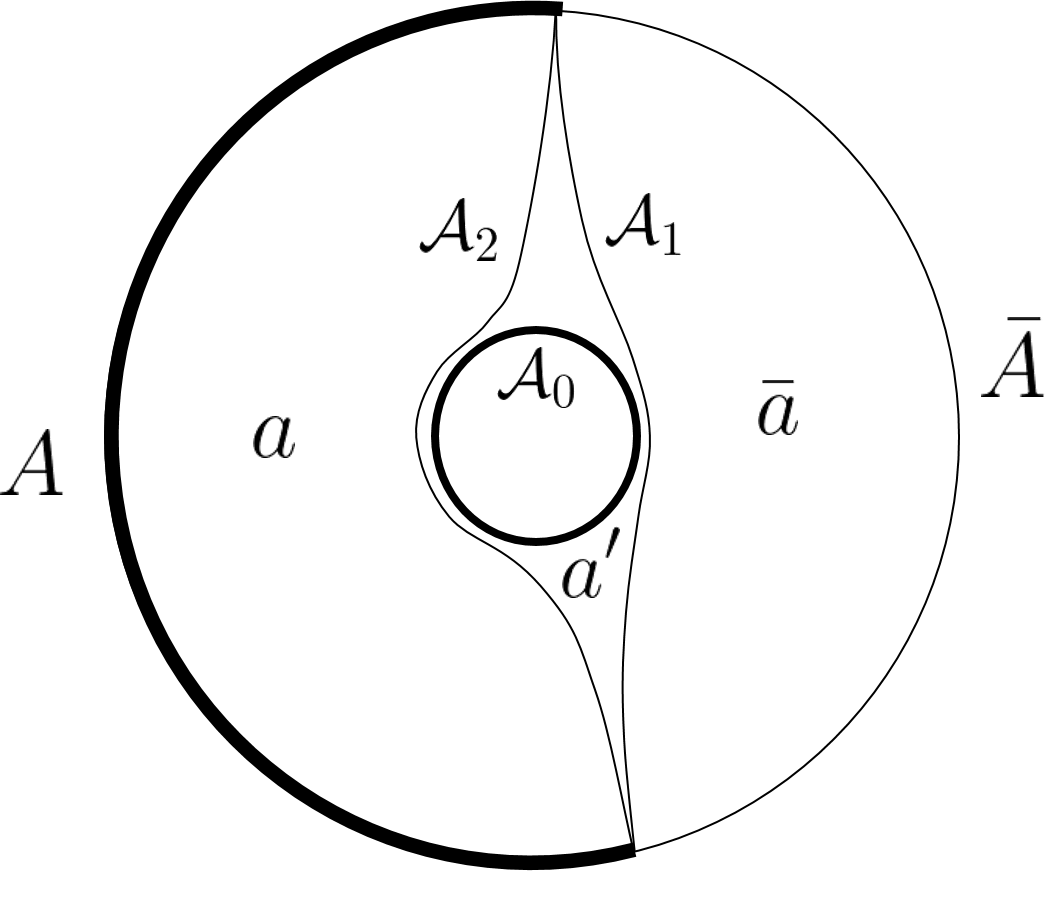}
 \centering
 \caption{A black hole in AdS}
 \label{fig:bhentwedge}
\end{subfigure}
\caption{Ryu-Takayanagi surfaces for empty AdS-space and a black hole in AdS. In each case, the boundary is separated into two regions, $A$ and $\bar{A}$. In empty AdS, the boundary between these regions is also the boundary of a single minimal surface through the bulk, with (divergent) area $\mathcal{A}$, which is known as the Ryu-Takayanagi or RT surface. This minimal surface separates the bulk into two regions, $a$ and $\bar{a}$, which we refer to as the entanglement wedges of $A$ and $\bar{A}$ respectively. When a black hole with horizon area $\mathcal{A}_0$ is introduced to the bulk, it creates infinitely many locally minimal surfaces with the same boundary as $A$ and $\bar A$. In particular we are interested in the minimal surface homologous to $\bar A$ with area $\mathcal{A}_1$ and the minimal surface homologous to $A$ with area $\mathcal{A}_2$. They divide the bulk into three regions $a$, $\bar{a}$ and $a'$, where $a'$ lies between the minimal surfaces and contains the black hole. For the thermal or canonical ensemble, regions $a$ and $\bar a$ form the entanglement wedges of $A$ and $\bar A$ respectively.}
\label{fig:entwedge}
\end{figure}

Let us split the boundary into two complementary regions $A$ and $\bar A$. Each boundary region has an associated entanglement wedge in the bulk that we label $a$ and $\bar a$, as shown in Figure \ref{fig:entwedge}. However both the area term (at subleading order in $G_N$) and the bulk entropy term of \eqref{eq:rtformula} will in general depend on the state of the system; hence both the entanglement wedges may be state-dependent. If the boundary state is pure, the two entanglement wedges, $a$ and $\bar a$ will be complementary bulk regions; their union will contain the entire bulk. However, in the case of mixed states, as well as states that are entangled with a second system, there may a non-empty third region $a'$ contained in neither entanglement wedge; one-sided (thermal) and two-sided black holes are respectively good examples of these two cases. 

Every bulk region has an associated von Neumann algebra acting on the bulk (code) Hilbert space; similarly boundary regions are associated with von Neumann algebras acting on the larger boundary Hilbert space. For both bulk and boundary, the algebra of a region forms the commutant of the algebra associated to its complementary region.

Since we are only interested in a single bulk geometry, there are no non-trivial operators in the centres of either the bulk or boundary algebras that are particularly relevant for our purposes.  Therefore, for pedagogical reasons, we shall mostly assume that the centres of all the algebras are trivial and hence that we can associate a subsystem Hilbert space to each region. This assumption, although incorrect, is commonly used in the literature for simplicity and clarity. The von Neumann algebra associated to each region is simply the algebra of operators acting on the associated subsystem. The isometric embedding of the code subspace into the larger CFT Hilbert space has the form
\begin{align}
\mathcal{H}_{\text{code}} \cong \mathcal{H}_a \otimes \mathcal{H}_{\bar{a}}\, (\otimes\, \mathcal{H}_{a'}) \subseteq \mathcal{H}_A \otimes \mathcal{H}_{\bar{A}} \cong \mathcal{H}_{\text{CFT}},
\end{align}
where $\mathcal{H}_{a'}$ appears if region $a'$ is non-empty. For convenience we will sometimes use $J: \mathcal{H}_{\text{code}} \to \mathcal{H}_{\text{CFT}}$ to represent the canonical embedding isometry. 

In Appendix \ref{sec:algebra}, we provide a more detailed description of the framework of operator algebra quantum error correction, which is necessary to talk about von Neumann algebras with non-trivial centres. This is particularly important for understanding code spaces with more than one semiclassical geometry, where the area of the RT surface corresponds to a non-trivial operator in the centre of the bulk algebras. All statements made in this section can be translated into statements about operator algebras (thus eliminating the incorrect assumption about the algebras' centres) simply by replacing tensor product factors with von Neumann algebras and their commutants, and replacing partial traces with restrictions to subalgebras.

If we assume trivial centres and hence a tensor product factorisation, the entanglement wedge reconstruction conjecture can be phrased as follows: the channel $\mathcal{N} = \Tr_{\bar{A}} \left(J(\cdot)J^\dagger \right)$ is a subsystem error correcting code for $\mathcal{H}_a$.\footnote{Without the assumption of trivial centres, it states that the algebra associated to region $A$ forms an operator algebra quantum error correcting code for the algebra associated to region $a$, see Appendix \ref{sec:algebra}.} This means that there exists some decoding channel $\mathcal{D}: \mathcal{H}_A \to \mathcal{H}_a$ such that for all states $\rho \in S \left(\mathcal{H}_a \otimes \mathcal{H}_{\bar{a}}\, (\otimes \,\mathcal{H}_{a'})\right)$,
\begin{align}
\left(\mathcal{D} \circ \mathcal{N}\right) \rho \approx \rho_a.
\end{align} 
In other words, region $A$ contains all the information necessary to approximately reconstruct the reduced density matrix of the state for region $a$.\footnote{Since the entanglement wedge $a$ is in general state-dependent, we shall see in Section \ref{sec:approxent} that we really the intersection of the entanglement wedge $a$ for all states in the code space.}

More commonly, we tend to think about quantum field theory using the Heisenberg rather than Schr{\"o}dinger picture. The adjoint decoding channel $\mathcal{D}^\dagger: \mathcal{B}(\mathcal{H}_a) \to \mathcal{B}(\mathcal{H}_A)$ is a unital completely-positive superoperator that maps bulk operators in $\mathcal{H}_a$ to boundary operators on $\mathcal{H}_A$. It is defined by
\begin{align} \label{eq:adjoint}
\Tr \left( \mathcal{D}^\dagger (X_a) \rho_A\right) = \Tr \left(X_a \mathcal{D} (\rho_A) \right),
\end{align}
for any observable $X_a \in \mathcal{B}(\mathcal{H}_a)$. We can therefore use $\mathcal{D}^\dagger$ to reconstruct operators in the bulk using operators in only a subregion of the boundary, so long as the bulk operators are contained in the entanglement wedge of the boundary subregion; the existence of such a map is the most commonly-used definition of entanglement wedge reconstruction.

\subsection{Entanglement wedge reconstruction from approximate decoupling} \label{sec:approxent}
Before considering the specific task of bulk reconstruction in black hole geometries, we first establish some more general facts about entanglement wedge reconstruction. It was argued in \cite{dong2016reconstruction} that a boundary region $A$ can be used to reconstruct a bulk region $a_0$ if the bulk region $a_0$ was not contained in the entanglement wedge of the complementary boundary region $\bar A$. Equivalently, region $a_0$ must be contained in the entanglement wedge of region $A$ for all \emph{pure} states.\footnote{We continue to distinguish the entanglement wedge $a$ from the decodable region $a_0$ because in general the entanglement wedge $a$ may depend on the state we are considering. Note that state-dependent entanglement wedges were not explicitly considered in \cite{dong2016reconstruction}. However, applying the reconstruction theorem from \cite{dong2016reconstruction} to code spaces with state-dependent entanglement wedges and ignoring approximation issues, one finds that the decodable region $a_0$ must be contained in the entanglement wedge $a$ for \emph{all} pure states in $\mathcal{H}_{\text{code}}$.} 

For simplicity, the technical arguments in \cite{dong2016reconstruction} ignored the existence of finite $G_N$ corrections that make any reconstruction at best approximate; however the authors expected that the arguments should generalise to the approximate setting. For the situations considered in \cite{dong2016reconstruction}, where the code space dimension remains fixed in the limit $G_N \to 0$, this is indeed the case, as was verified in \cite{cotler2017entanglement}.

However, while the technical results of \cite{dong2016reconstruction} continue to be true for exact error correcting codes, even when the code space dimension is very large (e.g. when the code space contains many black hole microstates), the generalisation to approximate error correction does not. 

Recall that we saw in Section \ref{sec:abits} that being able to universally and exactly decode small subspaces implies being able to exactly decode the entire space, but that the approximate version of this statement was not true, because the quality of the approximation could degrade linearly with the dimension of the decoded space. Furthermore, to obtain dimension-independent bounds on the decoding error, we had to consider the forgetfulness of the environment for input states entangled with a reference system of equal dimension to the subspaces being decoded. No reference system was necessary when the error correction was exact.

The same phenomenon will appear when we try to prove approximate entanglement wedge reconstruction. It is not enough for the entanglement wedge of $\bar A$ to never contain the bulk region $a_0$; instead the entanglement wedge of $\bar A \cup R$ cannot contain the bulk region $a_0$ for any (potentially entangled) pure state $\ket{\psi} \in \mathcal{H}_{\text{code}} \otimes \mathcal{H}_R$, where $R$ is a reference system with the same dimension as the code space $\mathcal{H}_{\text{code}}$. Equivalently, the entanglement wedge $a$ of $A$ needs to always contain the bulk region $a_0$ for any pure, \emph{or mixed} state.  Hence region $a_0$ is really the intersection of the entanglement wedges of $A$ for all pure states $\ket{\psi} \in \mathcal{H}_{\text{code}} \otimes \mathcal{H}_R$.) The weaker Dong-Harlow-Wall condition that a bulk operator only needs to be contained in the entanglement wedge $a$ for all pure states is only sufficient to show that the \emph{zero-bits} of the operator can be reconstructed in region $A$. 

Our justification for this claim involves a significant amount of technology from quantum information and takes the rest of this section: only the basic conclusions already mentioned will be required to understand the essential arguments in Section \ref{sec:BH}. The same results were also reached via very different arguments in \cite{cotler2017entanglement}. We choose to make a decoupling-based argument here, firstly because it makes explicit exactly when and how the Dong-Harlow-Wall argument from \cite{dong2016reconstruction} fails and secondly because we can use it to prove that reconstruction can be made exact to all orders in perturbation theory. It appears significantly harder to adapt the argument in \cite{cotler2017entanglement} to be perturbatively exact.

A version of the argument used by Dong, Harlow and Wall in \cite{dong2016reconstruction} to justify the entanglement wedge reconstruction conjecture starts as follows. It had been shown in \cite{jafferis2016relative} that in the limit $N \to \infty$ relative entropies in the bulk become equal to relative entropies on the boundary. This means that if any two states, $\rho$ and $\sigma$ in the code subspace satisfy
\begin{align} \label{eq:rhoeqsigma}
\rho_{\bar a} = \sigma_{\bar a},
\end{align}
then 
\begin{align} \label{eq:relent}
S(\rho_{\bar A} || \sigma_{\bar A}) \leq \varepsilon
\end{align}
for some small $\varepsilon$ that tends to zero if $N \to \infty$.  In general the difference between the bulk and boundary relative entropies will be $O(G_N)$. However, $\rho_{\bar a} = \sigma_{\bar a}$ implies that the quantum extremal surface should be the same for the two states $\rho, \sigma$ to all orders in perturbation theory. Hence $\varepsilon$ will be non-perturbatively small (although still non-zero) for small $G_N$ \cite{dong2018entropy}. 

If  $\varepsilon = 0$, then for all pairs of states $\rho, \sigma \in S(\mathcal{H}_{\text{code}})$ satisfying \eqref{eq:rhoeqsigma}, it can be shown that there must exist a channel $\mathcal{D}: S(\mathcal{H}_A) \to S(\mathcal{H}_a)$ such that
\begin{align}
\mathcal{D} \circ \Tr_{\bar{A}}(\rho) = \rho_a,
\end{align}
for all states $\rho \in S(\mathcal{H}_{\text{code}})$. (In fact it would imply that we could additionally reconstruct region $a'$ if such a region existed for some mixed states.) The channel $\Tr_{\bar A}(J(\cdot)J^\dagger)$ would therefore form an exact error-correcting subsystem code for $\mathcal{H}_a$, which is what we wanted to show. This was the main technical result of \cite{dong2016reconstruction}.

While acknowledging that the equality between bulk and boundary relative entropy, and hence entanglement wedge reconstruction, should only be approximate at finite $G_N$, \cite{dong2016reconstruction} did not attempt to prove an approximate version of their decoupling theorem, and instead left it as a task for future work. Let us now attempt to do exactly that. To generalise the argument given above to apply even for small but non-zero $\varepsilon$ requires a generalisation of (\ref{eq:forgetful}), or (equivalently) a version of Kretschmann \emph{et al.}'s information-disturbance theorem \cite{kretschmann2006information} that works for subsystem error-correcting codes. Fortunately, such a generalisation is relatively straightforward and was done for the even more general structure of operator algebra error correction in \cite{beny2009conditions}. We discuss the general form and its applicability in Appendix \ref{sec:algebra}, but for now we shall simply apply the result in the special case of subsystem error correction.  

As in Section \ref{sec:abits}, we only obtain dimension-independent bounds if we consider states that are entangled with a reference system $\mathcal{H}_R$ that has the same dimension as the space of states we wish to decode.

We first define an approximate subsystem error correcting code as follows. Let the channel $\mathcal{N}: S(\mathcal{H}_{a_0} \otimes \mathcal{H}_{\bar a_0}) \to S(\mathcal{H}_A)$. The channel $\mathcal{N}$ forms an approximate subsystem error correcting code with error $\delta_1$ if
\begin{align} \label{eq:epsilon}
\delta_1 = \inf_{\mathcal{D}} \left\lVert \mathcal{D} \circ \mathcal{N} - \Tr_{\bar b} (\cdot)\right\rVert_\diamond.
\end{align}
where the infimum is taken over decoding channels $\mathcal{D}: S(\mathcal{H}_A) \to S(\mathcal{H}_b)$. 

The complementary channel $\mathcal{N}^c: $ completely forgets the subsystem $\mathcal{H}_{a_0}$ with uncertainty $\delta_2$ if
\begin{align} \label{eq:delta}
\delta_2 = \sup_{\ket{\psi}}\left\lVert \psi_{\bar A R}  - \Tr_A \left( \omega_{a_0} \otimes \psi_{\bar a_0 R} \right)\right\rVert_1
\end{align}
where $\omega_{a_0} \in S(\mathcal{H}_{a_0})$ is maximally mixed and the supremum is again over all states $\ket{\psi} \in \mathcal{H}_{a_0} \otimes \mathcal{H}_{\bar a_0} \otimes \mathcal{H}_R$. The dimension $d_R$ of the reference system can be unrestricted; however, it is sufficient to consider a reference system whose dimension is equal to the dimension of $\mathcal{H}_{\text{code}}$.

What does \eqref{eq:delta} mean in the context of entanglement wedge reconstruction? Let $a_0$ be the intersection of the entanglement wedges of $a$ for all states $\ket{\psi} \in \mathcal{H}_{\text{code}} \otimes \mathcal{H}_R$ and let $\bar a_0$ be its bulk complement. Hence we have
\begin{align}
\mathcal{H}_{\text{code}} \cong \mathcal{H}_{a_0} \otimes \mathcal{H}_{\bar a_0}.
\end{align}
If, as before, we have $\mathcal{N} = \Tr_{\bar A}(J(\cdot)J^\dagger)$, then the complementary channel $\mathcal{N}^c = \Tr_A(J(\cdot)J^\dagger)$. We note that because in \eqref{eq:delta} we are considering pure states $\ket{\psi} \in \mathcal{H}_{A} \otimes \mathcal{H}_{\bar A} \otimes\mathcal{H}_R$, the entanglement wedge of $\bar A \cup R$ is by definition the complement of the entanglement wedge of $A$. In other words, the entanglement wedge of $\bar A \cup R$ is given by $a' \cup \bar{a} \cup R$.\footnote{Some readers may be unhappy at the notion of an entanglement wedge for states entangled with a reference system, which is not itself holographic. If so, it may be comforting to imagine the reference system as a second copy of the CFT with the same bulk code subspace so that everything is holographic. Note that, depending on context, $R$ then refers to either the entire boundary or the entire bulk of this second system. We also remind readers that the arguments in \cite{cotler2017entanglement} give the exact same conclusion we reach here (region $a_0$ must be contained in the entanglement wedge of $A$ for all pure \emph{or mixed} states) without ever invoking a reference system. We only need to do so here to make comparisons with the arguments in \cite{dong2016reconstruction}.} By definition this is contained in $\bar a_0 \cup R$. Therefore
\begin{align}
\left( \omega_{a_0} \otimes \psi_{\bar a_0 R} \right)|_{a'\bar{a}R} = \psi_{a' \bar a R}
\end{align}
and, hence, by the approximate equality between bulk and boundary relative entropies \cite{jafferis2016relative, dong2018entropy}, we have
\begin{align}
S(\Tr_A \left( \omega_{a_0} \otimes \psi_{\bar a_0 R} \right) || \psi_{\bar A R}) \leq \varepsilon,
\end{align}
for some non-perturbatively small $\varepsilon$. Using Pinsker's inequality \cite{hiai2008sufficiency}, it follows that $\delta_2 \leq \sqrt{2\,\varepsilon\, \ln 2 }$ and so this will also be non-perturbatively small.

However the size of the uncorrectable error $\delta_1$ and the uncertainty in the forgetfulness $\delta_2$ are related by \cite{beny2009conditions}
\begin{align} \label{eq:deltaepsilon}
\frac{1}{4} \delta_2^2 \leq \delta_1 \leq 2 \delta_2^{\frac{1}{2}},
\end{align}
and hence both tend to zero simultaneously with dimension-independent bounds. (The equivalent result for general operator algebra error correction is reproduced in Appendix \ref{sec:algebra} as Theorem \ref{thrm:alinfdist}; the subsystem error correction case \eqref{eq:deltaepsilon} follows as a trivial consequence from this.) We have therefore shown that there exists an approximate subsystem error correcting code for region $a_0$ with non-perturbatively small error.

As for ordinary subspace error correction (discussed in Section \ref{sec:abits}), such dimension-independent bounds are not possible if we do not include a reference system with the same dimension as the code space in the definition of complete forgetfulness. If  the dimension of the code space is fixed in the semiclassical limit $G_N \to 0$, this is not especially problematic. It may contribute a large constant factor to the size of the decoding error, but it cannot affect with the error tends to zero in the limit $G_N \to 0$. This will not be true if want our code space $S$ to include a large number of black hole microstates. Since the code space dimension $d_{\text{code}} \to \infty$ if $G_N \to 0$, dimension-dependent factors can affect whether the error tends to zero in this limit. As a result, they cannot be safely ignored. 

If the code space dimension is fixed, the RT formula is dominated by the classical area term in the semiclassical limit and the RT surface is state-independent.  However if the code space dimension diverges sufficiently fast, the bulk entropy term can compete with the area term and it matters whether we consider the entanglement wedge of mixed states or only pure states. It is not a conincidence that a divergent code space dimension is required for both; the Dong-Harlow-Wall argument gives qualitatively different conclusions to the conclusions derived here --  in exactly the contexts where we have shown that its conclusions should not be trusted.

\subsection{Entanglement wedges for code spaces containing black holes} \label{sec:BH}
We argue in Section \ref{sec:microspace}, that we can construct code spaces $$\mathcal{H}_{\text{code}} \cong \mathcal{H}_{\text{BH}} \otimes \mathcal{H}_{\text{ext}},$$
where the dimensions $d_{\text{code}}$ and $d_{\text{BH}}$ satisfy
$$ \lim_{G_N \to 0} 4 G_N \log d_{\text{BH}} = \lim_{G_N \to 0} 4 G_N \log d_{\text{code}} = \mathcal{A}_0,$$
for a black hole with horizon area $\mathcal{A}_0$. The Hilbert space $\mathcal{H}_{\text{ext}}$ describes the degrees of freedom outside the horizon, while $\mathcal{H}_{\text{BH}}$ describes the microstate of the black hole itself. Bulk operators outside the black hole horizon act only $\mathcal{H}_{\text{ext}}$, while the degrees of freedom in $\mathcal{H}_{\text{BH}}$ are localised to the black hole (in other words the state on a boundary region only depends on $\mathcal{H}_{BH}$ if the entanglement wedge of the boundary region contains the black hole). Moreover, all of the microstates in this code space are thermalised, typical ``equilibrium states''.\footnote{This last property is only possible because we are not trying to include \emph{all} black hole microstates in our code space, merely sufficiently many microstates to give the Bekenstein-Hawking entropy up to a subleading correction.}

In Figure \ref{fig:bhentwedge}, we show an area $A$ of the boundary together with two extremal surfaces through the bulk whose boundary is $\partial A$.  The first is homologous to $\bar{A}$ (the black hole is between the minimal surface and $A$) and has area $\mathcal{A}_1$. The second is homologous to $A$ (the black hole is between the minimal surface and $\bar{A}$) and has area $\mathcal{A}_2$. We shall assume $\mathcal{A}_1 < \mathcal{A}_2 < \mathcal{A}_1 + \mathcal{A}_0$. We label the bulk region between $A$ and the minimal surface with area $\mathcal{A}_2$ by $a$, while the region between $\bar{A}$ and the minimal surface with area $\mathcal{A}_1$ is called $\bar{a}$ and the region between the two minimal surfaces (which contains the black hole) is labelled $a'$. Note that, for the thermal state, region $a$ is the entanglement wedge of region $A$ and region $\bar a$ is the entanglement wedge of region $\bar a$ and so this is consistent with our previous notation. However, for individual microstates, the entanglement wedge of $A$ will be region $a \cup a'$. To avoid confusion, we will keep the definitions of regions $a$, $a'$ and $\bar a$ fixed and state-independent throughout this section, rather than having their definition depend on the state.

The code space has the form $$\mathcal{H}_{\text{code}} \cong \mathcal{H}_a \otimes \mathcal{H}_{a'} \otimes \mathcal{H}_{\bar a}.$$ Since the black hole is contained within region $a'$, we have
$$\mathcal{H}_{a'} \cong \mathcal{H}_{\text{BH}} \otimes \mathcal{H}_{a'}^{\text{ext}}.$$
Since we chose the code space geometry to include only a single black hole plus perturbative excitations outside it, only $\mathcal{H}_{a'}$ (and not $\mathcal{H}_a$ or $\mathcal{H}_{\bar a}$)  contains an exponential number of states (w.r.t. $1/G_N$).

Now consider an arbitary pure state $\ket{\psi} \in \mathcal{H}_{\text{code}} \otimes \mathcal{H}_R$. We wish to find the entanglement wedge of $\bar{A} \cup R$ for all such states. If some bulk region $a'$ is not contained in the entanglement wedge of $\bar A \cup R$ for any state in $\mathcal{H}_{\text{code}} \otimes \mathcal{H}_R$, then, by our arguments in Section \ref{sec:approxent}, it should be possible to reduced density matrices for region $a \cup a'$ from reduced density matrices for region $\bar A$. Equivalently it should be possible to simulate operators acting on $\mathcal{H}_a \otimes \mathcal{H}_{a'}$ with operators acting on $\mathcal{H}_A$.

\begin{figure}[t]
\includegraphics[width = 0.6\linewidth]{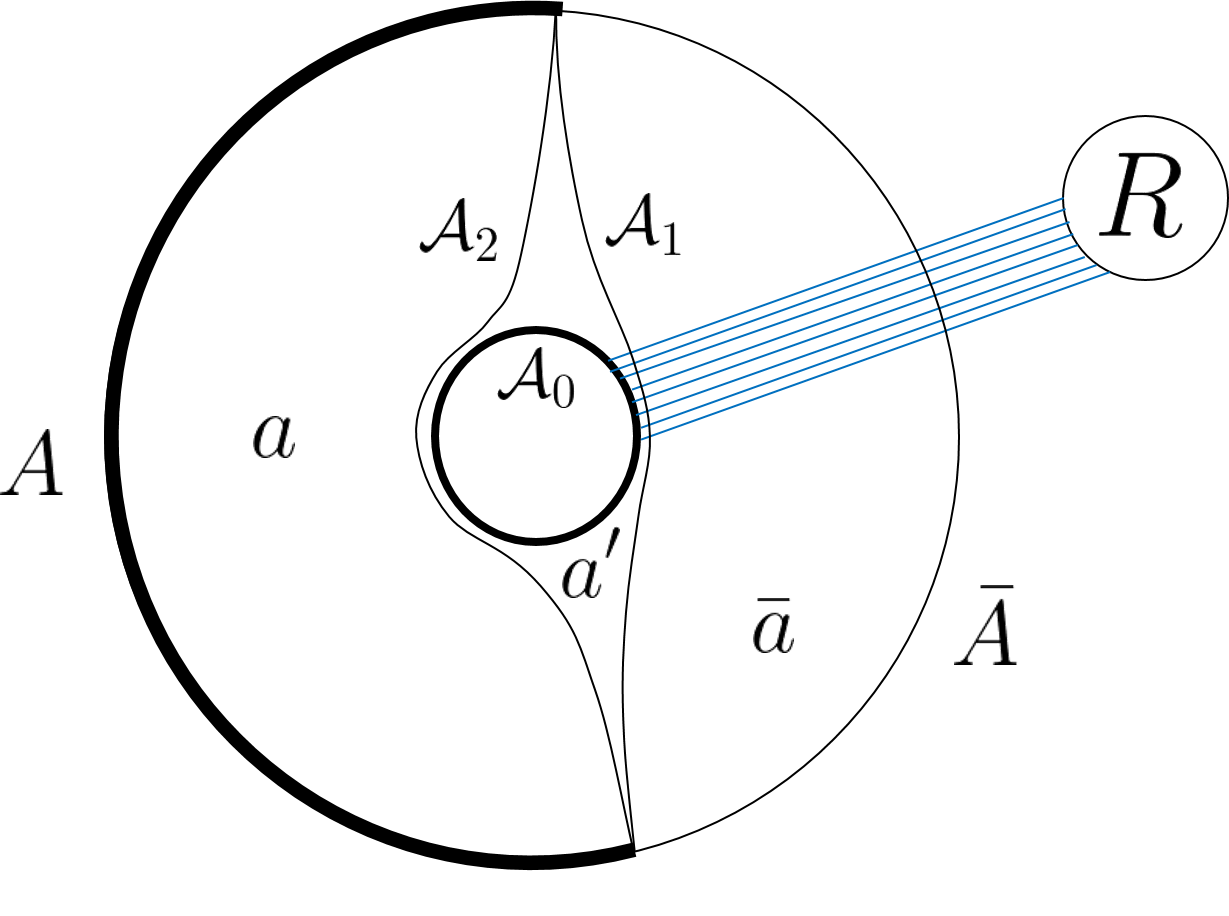}
\centering
\caption{The black hole is now entangled with a reference system $R$. If the entanglement between the black hole and $R$ has entropy greater than $\frac{\mathcal{A}_2 - \mathcal{A}_1}{4 G_N}$, the entanglement wedge of $\bar{A} \cup R$ is $\bar{a} \cup a' \cup R$. Otherwise the entanglement wedge of $\bar{A} \cup R$ is only $\bar{a} \cup R$.}
\label{fig:bhrentwedge}
\end{figure}

What can this entanglement wedge be? Since we chose the code space such that the geometry is approximately the same for every state in the subspace, the first term in the Ryu-Takayanagi formula (\ref{eq:rtformula}) is approximately the same for all code states and any fixed surface. As a result the only way state dependence of the entanglement wedge can arise is through the second term in (\ref{eq:rtformula}).\footnote{In principle, as with the thermofield double state, there could be a smooth geometry between the horizon of the black hole and $R$ with an associated  minimal area; however, we are always free to interpret this as simply bulk entanglement, see \cite{harlow2017ryu}.} Moreover, the only source of bulk entanglement that can be sufficiently large to compete with the area term as $G_N \to 0$ is entanglement between the black hole and reference system. As a result, the quantum minimal surface will always be one of the two classical extremal surfaces with areas $\mathcal{A}_1$ and $\mathcal{A}_2$. The only question is whether the entanglement wedge of $\bar{A} \cup R$ contains only $\bar{a} \cup R$ or consists of $a' \cup \bar{a} \cup R$, as depicted in Figure \ref{fig:bhrentwedge}. We can therefore decode $a \cup a'$ (and hence the black hole) from $A$ so long as for all states $\ket{\psi} \in \mathcal{H}_{\text{code}} \otimes \mathcal{H}_R$,
\begin{align} \label{eq:comp}
S(\bar{a}a'R) _\psi + \frac{\mathcal{A}_2}{4G_N} > S(\bar{a}R)_\psi + \frac{\mathcal{A}_1}{4G_N},
\end{align}
We have $S(a'\bar{a}R)_\psi = S(a)_\psi = O(1)$ and, by the triangle inequality, $\left| S(R)_\psi - S(\bar{a}R)_\psi \right| \leq S(\bar{a}) = O(1)$. Hence, to leading order in $G_N$, (\ref{eq:comp}) is equivalent to
\begin{align} \label{eq:breakentwedge}
4G_N S(R)_\psi < \mathcal{A}_2 - \mathcal{A}_1,
\end{align}
If 
\begin{align}
\lim_{G_N \to 0} 4 G_N \log d_R = \lim_{G_N \to 0} 4 G_N \log d_S  < \mathcal{A}_2 - \mathcal{A}_1,
\end{align}
then this will be satisfied for any state $\ket{\psi} \in \mathcal{H}_{\text{code}} \otimes \mathcal{H}_R$. Conversely if 
\begin{align}
\lim_{G_N \to 0} 4 G_N \log d_R = \lim_{G_N \to 0} 4 G_N \log d_S  > \mathcal{A}_2 - \mathcal{A}_1,
\end{align}
then it will be violated for any maximally-entangled state on $\mathcal{H}_{\text{code}} \otimes \mathcal{H}_R$. We therefore see that in the classical limit we can decode $\mathcal{H}_a \otimes \mathcal{H}_a'$ from $\mathcal{H}_A$ for any subspace $\mathcal{H}_S \subseteq \mathcal{H}_{\text{code}}$ whose dimension is less than $e^{\alpha \frac{\mathcal{A}_0}{4G_N}}$ for
\begin{align} \label{eq:alphaentwedge}
\alpha = \frac{\mathcal{A}_2 - \mathcal{A}_1}{\mathcal{A}_0}.
\end{align}
In other words $A$ contains the $\alpha$-bits of region $a'$ for the entire code space $\mathcal{H}_{\text{code}}$. This region contains not just the black hole, but also an additional bulk region outside the black hole horizon but between the two minimal surfaces. In contrast the region $a$ can be decoded for the entire code space, since this region will always lie outside the entanglement wedge of region $\bar A \cup R$.

In the Heisenberg picture, this means that we can simulate an operator $\mathcal{O}_{a'}$ on $a'$ with an operator  $\mathcal{O}_{A}$ on $A$ so long as we only require that the operator $\mathcal{O}_{A}$ behave in the same way as $\mathcal{O}_{a'}$ within a subspace $\mathcal{H}_S \subseteq \mathcal{H}_{\text{code}}$ with dimension less than $e^{\alpha \frac{\mathcal{A}_0}{4G_N}}$. In other words, for any bulk operator in $a'$, whether acting on the black hole or outside the horizon, the operator on $A$ must be state-dependent. We discuss the connection with other proposed forms of operator state dependence in quantum gravity in Section \ref{sec:statedepend}.

Finally, we want to show that the value of $\alpha$ given in \eqref{eq:alphaentwedge} is optimal. In other words, that for any $\alpha' > \alpha$, the $\alpha'$-bits of region $a'$ are not encoded in region $A$. We have already shown that \eqref{eq:breakentwedge} can be violated when $d_R = e^{\alpha' S}$ for such $\alpha'$ and hence that there exist states in $\mathcal{H}_{\text{code}} \otimes \mathcal{H}_R$ for which the entanglement wedge of $\bar{A} \cup R$ contains region $a'$. Specifically we can consider a state where $\mathcal{H}_R$ is maximally entangled with $\mathcal{H}_{BH}$. Acting with bulk operators in region $a'$ but outside the black hole horizon cannot change the entanglement wedge for such a state, and so we can construct a small subspace of states $\mathcal{H}_S \subseteq \mathcal{H}_{\text{code}} \otimes \mathcal{H}_R$ with dimension $d_S$ which look identical in region $a$ and for which region $a'$ is never contained in the entanglement wedge of $A$. Since the dimension of $\mathcal{H}_S$ is small, even if we consider states entangled with a second reference system $\mathcal{H}_{R'}$, the entanglement wedge of $A \cup R'$ will still never contain region $a'$. 

It follows, by the arguments made in Section \ref{sec:approxent}, that we can recover the state for region $a'$ from $\bar A \cup R$ so long as we know that it lies in the subspace $\mathcal{H}_{S}$. By the no cloning theorem (or, more formally, Kretschmann \emph{et al.}'s information-disturbance theorem \cite{kretschmann2006information}), we therefore cannot recover the state from $A$. However the support of any state in $\mathcal{H}_S$ lies in a  subspace $\mathcal{H}_{\text{sup}} \subseteq \mathcal{H}_{\text{code}}$ of dimension at most $d_R \,d_S$. It therefore cannot be possible to decode region $a'$ using only region $A$ for the subspace $\mathcal{H}_{\text{sup}}$. Since $d_S$ is small and fixed, 
\begin{align}
\lim_{G_N \to 0} \frac{\log (d_R \,d_S)}{S_{BH}} = \alpha'.
\end{align}
Since we could have chosen $\alpha'$ to be arbitrarily close to $\alpha$, region $A$ cannot encode the $\alpha'$-bits of region $a'$ for any $\alpha' > \alpha$.

This argument makes clear that the operator state dependence is unavoidable; it is not simply the product of a particular reconstruction strategy. However, as we make region $A$ larger, not only does the region $a$ where operators can always be decoded become larger, so does the value of $\alpha$ itself. The size of the subspaces $\mathcal{H}_S$, for which operators in region $a'$ can be reconstructed, grows as region $A$ grows. Eventually, we reach the point where $\alpha = 1$ and hence a single operator will exist that exhibits the correct behaviour for all the black hole microstates in $\mathcal{H}_{\text{code}}$. The entanglement wedge of $A$ will now always contain regions $a$ and $a'$; for the purposes of entanglement wedge reconstruction, it no longer matters which of these regions an operator is in.

\section{Alpha-bits of BTZ black holes} \label{sec:btz}
We now consider the case of an uncharged, non-rotating BTZ black hole in $2+1$ dimensions. This provides a sufficiently simple example of the phenomena introduced in Section \ref{sec:entwedge} that many of the relevant quantities can be calculated analytically. Most of these calculations have already been done in the literature \cite{shenker2014black, bao2017distinguishability}. In particular the Holevo information $\chi$ for an ensemble of black hole microstates in AdS$_3$ was calculated in \cite{bao2017distinguishability}; it turns out that the Holevo information has a very simple relation to $\alpha$ which is given by\footnote{It should be obvious that $\alpha S$ is a lower bound for the Holevo information. The fact that \eqref{eq:holevo} is actually an equality, however, is a non-trivial fact about black holes and comes from the universal behaviour of black hole microstates.}
\begin{align} \label{eq:holevo}
\chi = \frac{\mathcal{A}_2 - \mathcal{A}_1}{4G_N} = \alpha S.
\end{align}
Nonetheless we shall carry out all the calculations here explicitly in the interest of clarity. 

In general, the explicit calculations conform with one's intuition; increasing the size of the boundary region $A$ increases $\alpha$, while increasing the radius of the black hole decreases $\alpha$. We also explicitly calculate the volume of the region $a'$ and find that it is equal to $2 \pi L^2$ -- independent of the radius of the black hole. The size of the region is always approximately AdS scale, even for very large black holes. The size is independent of $G_N$ -- $\alpha$-bit codes exist even in the semiclassical limit (in fact that is where they are best defined) -- but it cannot be made significantly larger than the AdS scale. The same effect is seen in Section \ref{sec:tensor} in tensor network toy models of holographic $\alpha$-bit codes. 

One possible explanation for this is that $\alpha$-bit codes in general seem to rely on the properties large random-like unitaries -- essentially they rely on and reflect some form of scrambling. However, fast scrambling only happens in large AdS black holes up to the AdS scale \cite{sekino2008fast}. More generally, locality above the AdS scale in AdS/CFT comes from locality in the CFT (and locality in energy scale for the radial dimension). However, sub-AdS scale locality is more mysterious and is associated with the large number of local matrix degrees of freedom in $\mathcal{N} = 4$ SYM at large $N$ (or the equivalent degrees of freedom in other examples).  

This suggests that $\alpha$-bit codes, or more specifically $\alpha$-bit degrees of freedom \emph{outside} the horizon, are a property of CFTs with large N (or large central charge). They are teaching us something about the encoding of degrees of freedom specifically for CFTs with weakly curved gravity duals. It would be interesting to see if the volume of region $a'$ continues to be AdS scale for more complex black holes, or whether, by tuning charges, angular momenta etc., we can make the region $a'$ arbitrarily large.

In Schwarzschild coordinates the BTZ metric is given by
\begin{align}
\text{ds}^2 = - \frac{r^2-R^2}{L^2} \text{dt}^2 + \frac{L^2}{r^2 - R^2} \text{dr}^2 + r^2 \text{d}\phi^2
\end{align}
where $\phi \sim \phi + 2 \pi$ and $R$ and $L$ are respectively the horizon radius and the AdS scale. 

Since gravity has no local degrees of freedom in $2+1$ dimensions, the BTZ black hole can be identified with a quotient of pure AdS$_3$. We can take advantage of this by calculating the area of minimal surfaces, which are just geodesics in $2+1$ dimensions, using formulas for the lengths of geodesics in pure AdS$_3$.
\begin{figure}[t]
\begin{subfigure}{.42\textwidth}
\includegraphics[width = .9\linewidth]{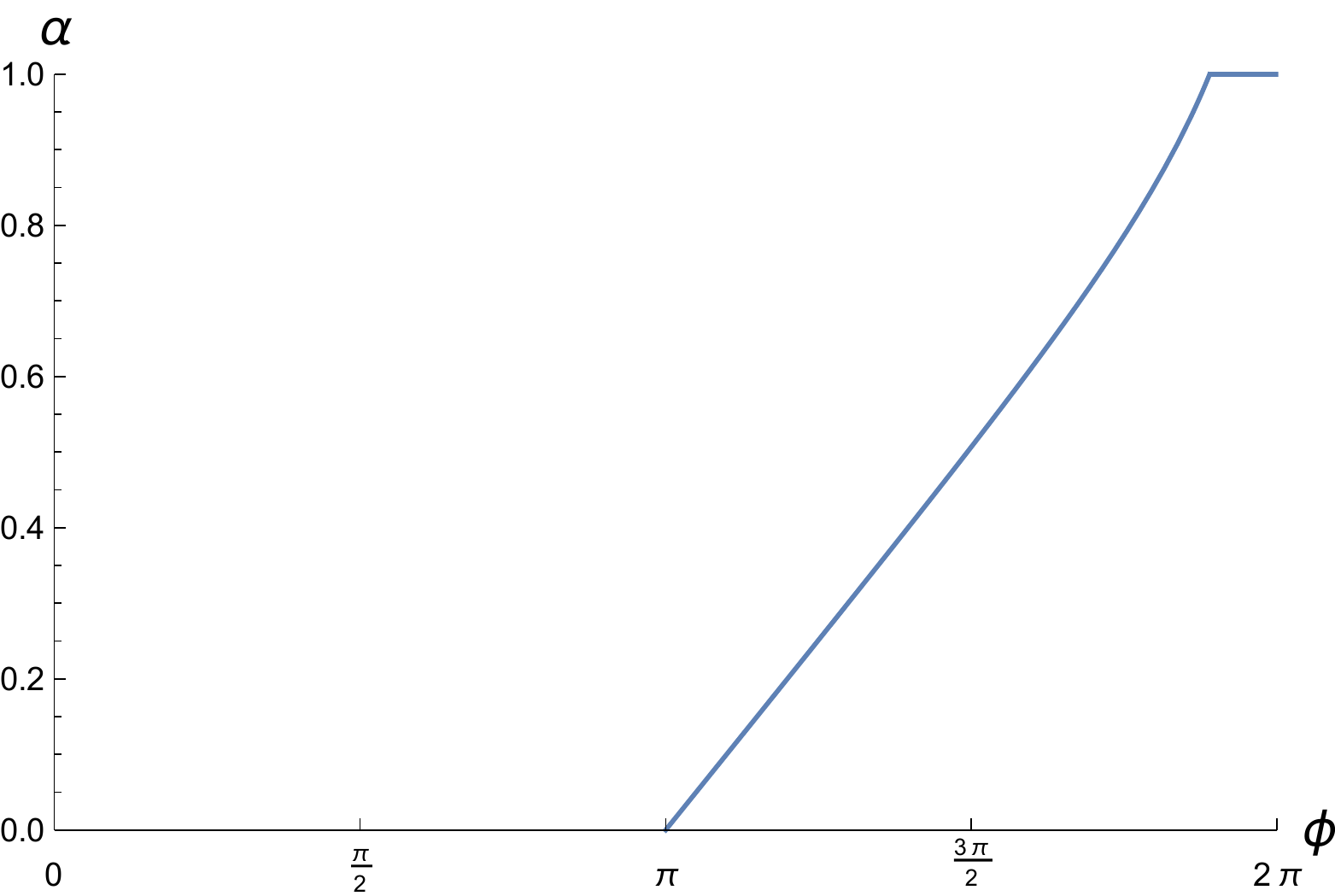}
\centering
\caption{Plot of $\alpha$ against boundary angle $\phi$.}
\label{fig:alpha}
\end{subfigure}
\begin{subfigure}{.55\textwidth}
\includegraphics[width = .9\linewidth]{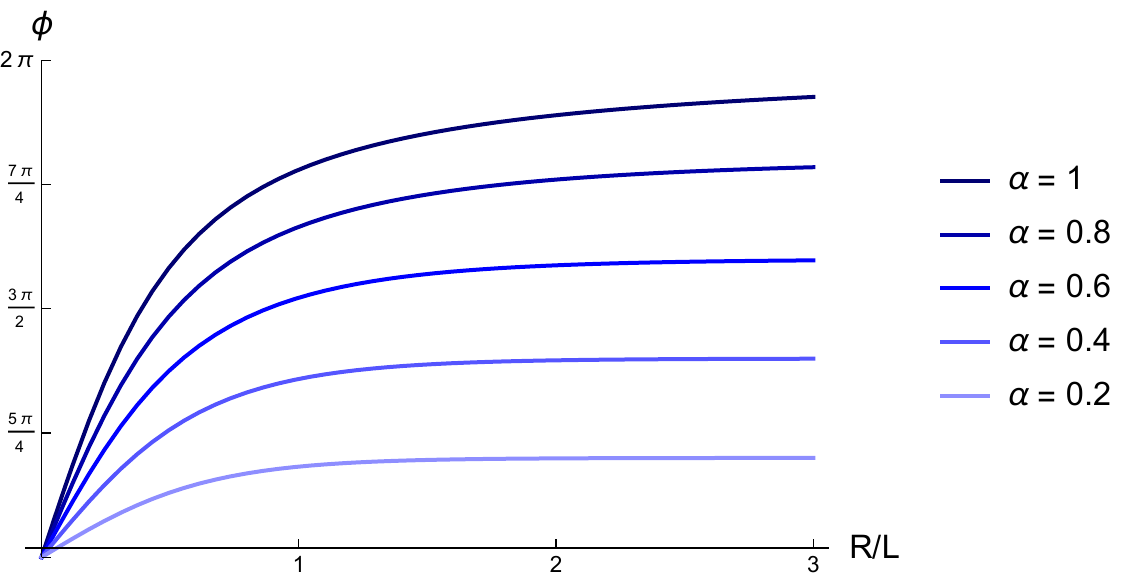}
\centering
\caption{Plot of angle $\phi$ against radius $R$ for various $\alpha$.}
\label{fig:radius}
\end{subfigure}
\caption{(a) $\alpha$ smoothly increases as we increase the size of the boundary region $A$ (with angle $\phi$). The black hole is a BTZ black hole with horizon radius $R = 2 L$. The boundary encodes no information about region $a'$ for $\phi < \pi$. For $\phi> \pi$, it contains the $\alpha$-bits for increasingly large $\alpha$, until eventually $\alpha = 1$ and the region $a'$ can be decoded without any knowledge of the black hole microstate. (b) The angle $\phi$ of the boundary region $A$ required to decode the bulk region $a'$ for fixed values of $\alpha$ steadily increases with the radius $R$ of the black hole in AdS units. As the radius becomes large, the required angle $\phi$ converges to $(1+\alpha)\pi$.}
\end{figure}

The geodesic length $d$ between two points in pure AdS satisfies
\begin{align} \label{eq:geo}
\text{cosh}\left(\frac{d}{l}\right) = T_1 T_1' + T_2 T_2' - X_1 X_1' - X_2 X_2'
\end{align}
where the embedding co-ordinates $(T_i,X_i)$ can be identified with Schwardschild co-ordinates $(r,t,\phi)$ by
\begin{align}
\begin{split}
&T_1 = \frac{1}{R} \sqrt{r^2 - R^2} \text{sinh} \frac{Rt}{L}, \\
&T_2 = \frac{r}{R} \text{cosh} \frac{R \phi}{L}, \\
&X_1 = \frac{1}{R} \sqrt{r^2 - R^2} \text{cosh} \frac{Rt}{L}, \\
&X_2 = \frac{r}{R} \text{sinh} \frac{R \phi}{L}.
\end{split}
\end{align}
The relevant geodesics on the BTZ geometry, which travel from $(r,0,0)$ to $(r,0,\phi)$ in the limit $r \to \infty$, can be identified with the geodesics in pure AdS from $(r,0,0)$ to $(r,0,\phi)$ and $(r,0,\phi-2\pi)$, since we identify $\phi \sim \phi + 2 \pi$ in the BTZ geometry. We label their lengths by $d_1$ and $d_2$ respectively. In the limit of large r, then (\ref{eq:geo}) implies,
\begin{align}
\frac{d_1}{L} = 2 \log \left( \frac{2r}{R} \right) + 2 \log \left(\text{sinh}\frac{R \phi}{2L}\right),
\end{align}
and
\begin{align}
\frac{d_2}{L} = 2 \log \left( \frac{2r}{R} \right) + 2 \log \left(\text{sinh}\frac{R (2 \pi -\phi)}{2L} \right).
\end{align}
Since the horizon area is simply $2\pi R$, we find that a boundary region of angle $\phi > \pi$ encodes the $\alpha$-bits of a BTZ black hole of horizon radius $R$ for
\begin{align}
\alpha = \frac{d_1 - d_2}{2\pi R} = \frac{\log \left(\text{sinh}\frac{R \phi}{2L} \right) -\log \left(\text{sinh}\frac{R (2 \pi - \phi)}{2L} \right)}{\pi \frac{R}{L}}.
\end{align}
Figure \ref{fig:alpha} shows how $\alpha$ increases smoothly with the size of the boundary region $A$ for a fixed black hole radius, while Figure \ref{fig:radius} shows how the size of the region required for any fixed $\alpha$ increases as the black hole radius increases.

If we take the limit where $\frac{R}{L} \gg 1$, while holding $\phi$ fixed, we find that
\begin{align}
\alpha = \frac{2\phi}{2 \pi} - 1.
\end{align}
The fraction of the boundary required to encode the $\alpha$-bits is the inverse of the $\alpha$-bit capacity of the noiseless qubit channel  $(1+\alpha)/2$. This is identical to the fraction of the Hawking radiation that we learned in Section \ref{sec:hawking} was required to encode the $\alpha$-bits of a black hole. We discuss this further in Section \ref{sec:abitsup}.

\begin{figure}[t]
\includegraphics[width = 0.4\linewidth]{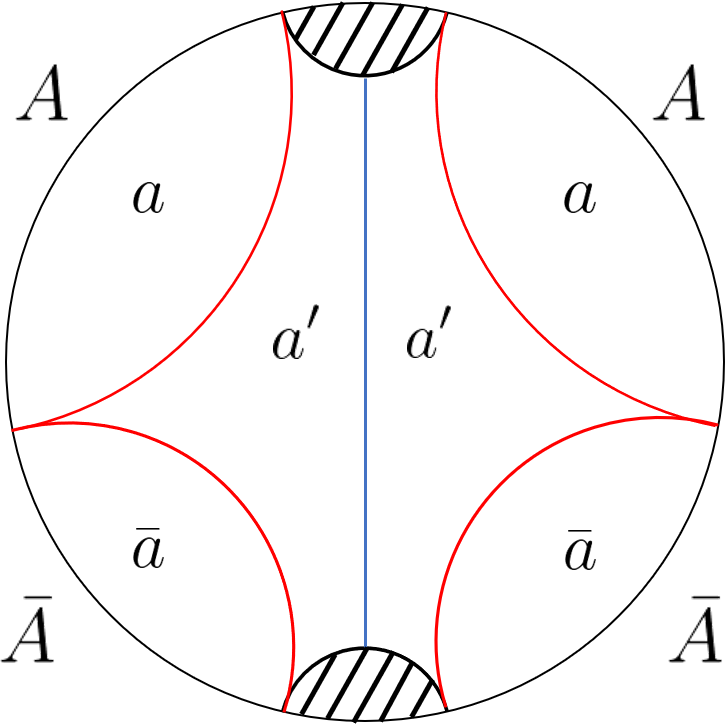}
\centering
\caption{A spatial slice of a two-sided BTZ black hole can be viewed as a quotient of the Poincar\'e disk. The fundamental domain is shown unshaded. We split each boundary into regions $A$ and $\bar A$, with an associated geodesic through the bulk (shown in red). The black hole horizon is shown in blue. Between the horizon and the geodesics (and within the fundamental region) there is a copy of region $a'$ on each side. The union of these two regions is bounded by an ideal hexagon on the Poincar\'e disk.}
\label{fig:btz}
\end{figure}
Finally we can calculate the size of the bulk region $a'$ (ignoring the black hole itself) for which the $\alpha$-bits are encoded in the boundary region $A$. An explicit calculation in Schwardschild co-ordinates seems daunting, so we shall instead again take advantage of the fact that the BTZ geometry is a quotient of pure AdS space. If we extend our picture of the BTZ black hole to include both sides of the Einstein-Rosen bridge, we see from Figure \ref{fig:btz} that two copies of region $a'$ forms the interior of a hexagon in hyperbolic space, bounded by two copies of each of the geodesics $A_1$ and $A_2$ together with the edges of the fundamental region of the BTZ quotient. 

This hexagon could be broken down in four triangles whose vertices all lie on the boundary of the space. These are known as ideal triangles and by using the symmetries of hyperbolic space, we can push the three boundary points to any other three boundary points: this shows that the area of the triangles (and hence $a'$) is independent of both the radius $R$ of the black hole and the boundary angle $\phi$.

To calculate the volume of the hexagon explicitly we can use the Gauss-Bonnet formula
\begin{align}
\int_{\mathcal M} \,K\, \text{dA} + \int_{\partial \mathcal{M}} k_g \text{ds} = 2 \pi \chi (\mathcal{M}).
\end{align}
The Gaussian curvature $K = -\frac{1}{L^2}$ and so can be taken out the front of the integral. Since the line segments are geodesics the only contribution to the boundary curvature term comes from the six corners, each of which lies on the boundary and so has angle $\theta \to \pi$. Finally the hexagon is homeomorphic to the disk and so has Euler characteristic $\chi(\mathcal M)= 1$. Evaluating this we obtain,
\begin{align}
V_{a'} = \frac{1}{2} V_{\text{hex}} = \frac{1}{2}\left( 6 \theta - 2 \pi\right) L^2 = 2 \pi L^2.
\end{align}
We see that the region $a'$ is always approximately AdS scale, as discussed above. A weakly curved bulk dual is required for region $a'$ to be sharply defined.

\section{Alpha-bits in tensor networks} \label{sec:tensor}
Tensor networks have been widely studied as a toy model of holography in recent years \cite{swingle2012entanglement,pastawski2015holographic,hayden2016holographic,czech2016tensor,verlinde2017emergent,osborne2017dynamics}. We find that they also give simple toy models of holographic $\alpha$-bit codes. Most importantly, they provide very clear intuition for why state dependence appears in the operator reconstruction; to reconstruct operators in the $\alpha$-bit region to the boundary we have to push them through the black hole itself. In other words, the isometry mapping bulk operators to boundary operators doesn't just depend on the state of the black hole; part of the isometry consists of the tensor that literally describes the state of the black hole.

We consider a version of the pentagon code developed in \cite{pastawski2015holographic}. Empty AdS in the bulk is described by perfect tensors, which have the property that they describe a unitary map from any half of their legs (or indices if one is more traditionally inclined) to the other half. Each six-legged tensor has a single dangling bulk leg. The overall network can be thought of as an isometry\footnote{If we view the tensor network a map from the bulk sites outwards to the boundary, we see that each tensor in the network has at least three legs flowing `outwards' and hence forms an isometry from the dangling and inwards pointing legs to the outwards legs. Additionally so long as $\log D < n_{\text{tot}} \log d$ the black hole random tensor will be approximately an isometry from the dangling microstates leg to the remaining legs. The entire network is therefore an isometry from bulk to boundary.} from a bulk Hilbert space to a large boundary Hilbert space. The image of the bulk Hilbert space can be thought of as the code subspace of the boundary space.

\begin{figure}[t]
\includegraphics[width = 0.45\linewidth]{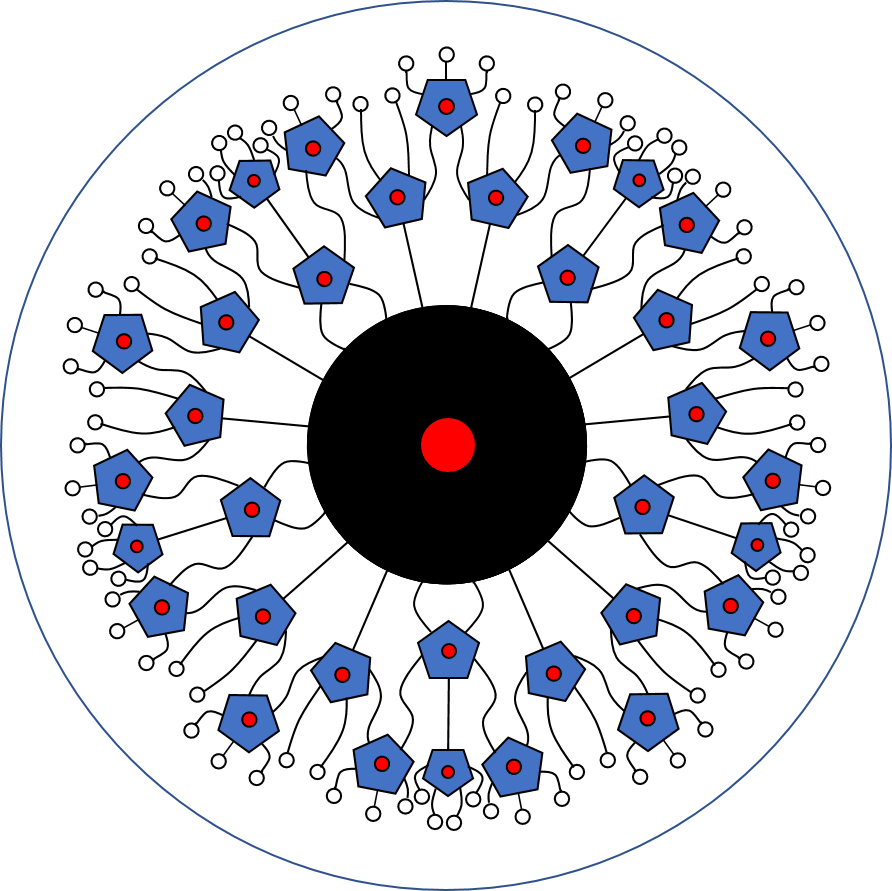}
\centering
\caption{A pentagon code with a black hole in the centre. The black hole is modelled as a large random tensor with a single $D$-dimensional dangling leg creating the space of allowed microstates and $n_{\text{tot}}$ $d$-dimensional legs flowing into the network. The remaining tensors are all perfect tensors with a single dangling $d$-dimensional leg that forms the bulk degrees of freedom and 5 $d$-dimensional legs flowing into the network.}
\label{fig:network}
\end{figure}
To include a black hole in this network, we simply add a random tensor with a large number of legs at the centre of the network, as shown in Figure \ref{fig:network}. If we wish to include an entire subspace of black hole microstate (rather than simply a single microstate), this tensor must have also have dangling bulk input leg whose dimension is the number of microstates we wish to consider.

If we divide the boundary into two regions $A$ and $\bar A$ as before, there is a natural notion of a bulk surface of `minimal area', which is simply the path through the bulk which cuts through the fewest bulk legs. If we have a black hole in the network, we can find the minimal surface on either side of the black hole. Just as in real AdS/CFT, there may exist tensors outside of the black hole that lie between these two geodesics, giving a region that naturally corresponds to the region $a'$ in Figure \ref{fig:bhentwedge}. Depending on the boundary points in question, the region $a'$ may contain anywhere from zero to two tensors adjacent to the black hole (as well as potentially other tensors further from the black hole). This is in agreement with our calculation in Section \ref{sec:btz}, where we found that volume outside of the black hole horizon for the region $a'$ in the BTZ geometry is of approximately AdS scale, independent of both the radius of the black hole and the angle of the boundary contained in region $A$.

Since the map from the bulk Hilbert space to the boundary Hilbert space is an isometry, clearly operators in the bulk can be represented (non-uniquely) by operators in the boundary. In fact we can use the properties of perfect tensors to do considerable better than this and represent operators in the bulk with operators on only part of the boundary; the isometry is an error-correcting code. To see why, note that we can view the perfect tensor at our bulk operator site as a unitary map from the bulk leg, as well as two other legs of our choice, to the remaining three legs. By conjugating the operator by this unitary, we can map it to an operator acting on only those three legs of the tensor network. In turn, we can `push' the operator acting on those three legs, through more tensors, so long as we always push the operator onto at least three new legs each time. In this way we can push the operator onto only a subregion of the boundary and achieve a version of entanglement wedge reconstruction.\footnote{In fact the region that can be reconstructed is slightly different to what one might naively guess, since the pentagon code does not always satisfy the Ryu-Takayanagi formula. Instead only the region known as the \textit{greedy entanglement wedge} can be reconstructed. For the same reason, the pentagon code will prove to be a weaker $\alpha$-bit encoding than true AdS/CFT.}
\begin{figure}[t]
\includegraphics[width = 0.6\linewidth]{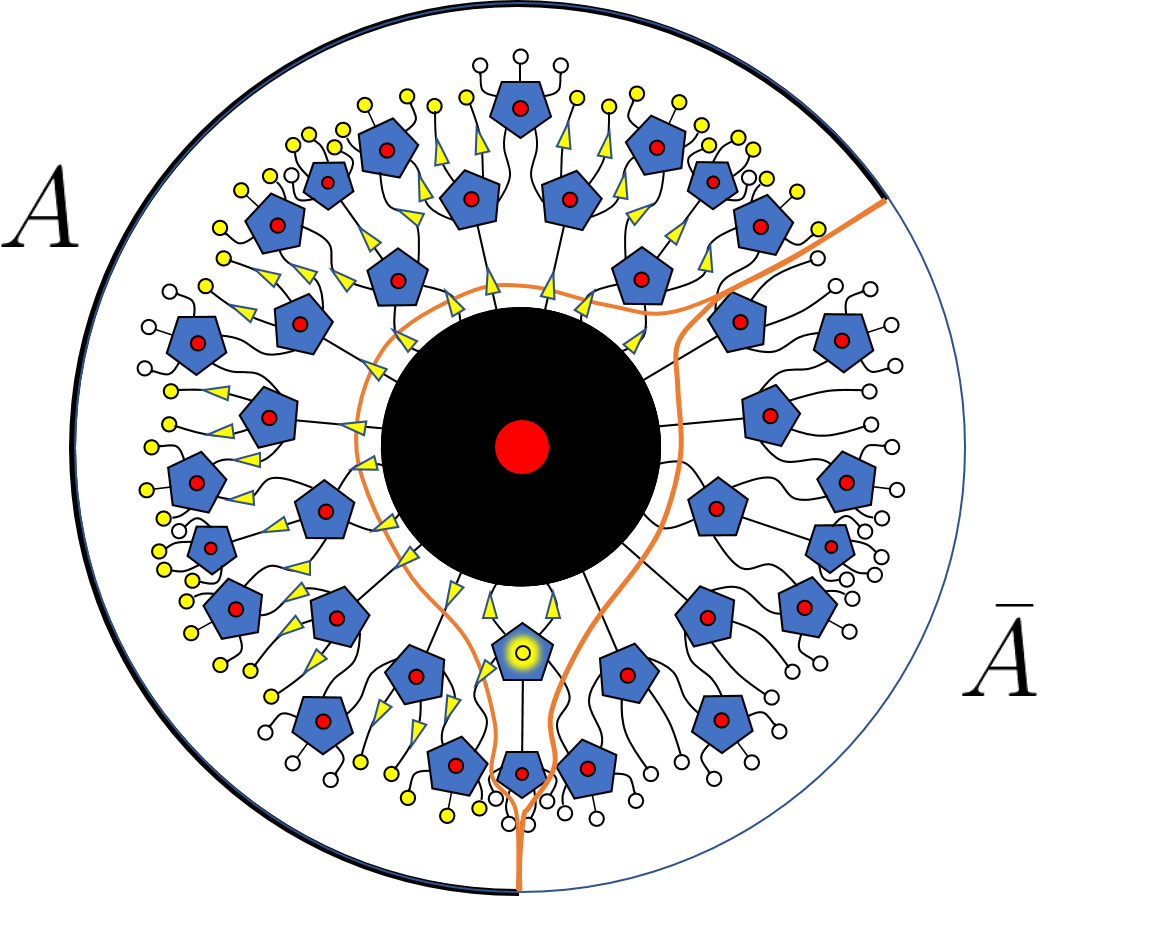}
\centering
\caption{An example of an operator that can only be pushed to region $A$ of the boundary through the black hole. This can only be achieved if the space of microstates is small enough for the black hole to act approximately as an isometry. In this case we have $n_a=12$, while $n_{\bar a} = 6$ and $n_{a'} = 2$. As a result we must have $\log D < 4\, \log d$, or equivalently $\alpha = \frac{1}{5}$.}
\label{fig:alpha-network}
\end{figure}

What happens if we have a situation equivalent to the one in Section \ref{sec:entwedge}? In this context that means that the difference between the size of the minimal cuts through the network on each side of the black hole is less than the number of legs coming out of the black hole. They naturally divide the bulk into regions $a$, $a'$ and $\bar a$, as in Figure \ref{fig:bhentwedge}. An example is shown in Figure \ref{fig:alpha-network}. In this case we see that it is impossible to push an operator in region $a'$ between the two cuts to the boundary without pushing it \textit{through} the black hole. Of course we can only push the operator through the black hole if the black hole tensor is (at least approximately) an isometry from the legs we treat as input to the legs we use as outputs.

A random tensor becomes approximately an isometry with very high precision in the limit of large dimension if the dimension of the input grows less quickly than the output dimension. If $n_a$ legs flow out of the black hole into region $a$, while $n_{a'}$ flow into region $a'$ and $n_{\bar a}$ legs flow into region $\bar a$, then, in the limit of large bond dimension $d$, we can approximately push the operator through the black hole (with its $D$ allowed microstates) and onto sites in region $a$ so long as
\begin{align}
\log D + \left(n_{\bar a}+ n_{a'}\right) \log d < n_{a} \log d.
\end{align}
In other words we require
\begin{align}
\log D < \left(n_{a}-n_{a'}-n_{\bar a}\right) \,\log d = \frac{n_{a}-n_{a'}- n_{\bar a}}{n_{\text{tot}}}  S
\end{align}
where $n_{\text{tot}}$ is the total number of legs flowing out of the black hole and $S = n_{\text{tot}}\,\log d$ is the entropy of the black hole. We can therefore consider at most $D = e^{\alpha S}$ microstates, where
$$\alpha = \frac{n_{a}-n_{a'}-n_{\bar a}}{n_{\text{tot}}}.$$
How does this compare to our formula for AdS/CFT? We have that
\begin{align}
\frac{A_2 -A_1}{A_0} = \frac{n_{a}+n_a^{\text{ext}}-n_{\bar a}-n_{\bar a}^{\text{ext}}}{n_{\text{tot}}},
\end{align}
where $n_a^{\text{ext}}$ and $n_{\bar a}^{\text{ext}}$ are the number of legs cut by the minimal surfaces which do not connect to the black hole. In the case of Figure \ref{fig:alpha-network}, 
$$n_a^{\text{ext}} = n_{\bar a}^{\text{ext}},$$
while $n_{a'}=2$, so we cannot decode as large subspaces in the pentagon code as we expect to be able to decode in AdS/CFT. 

\begin{figure}[t]
\includegraphics[width = 0.5\linewidth]{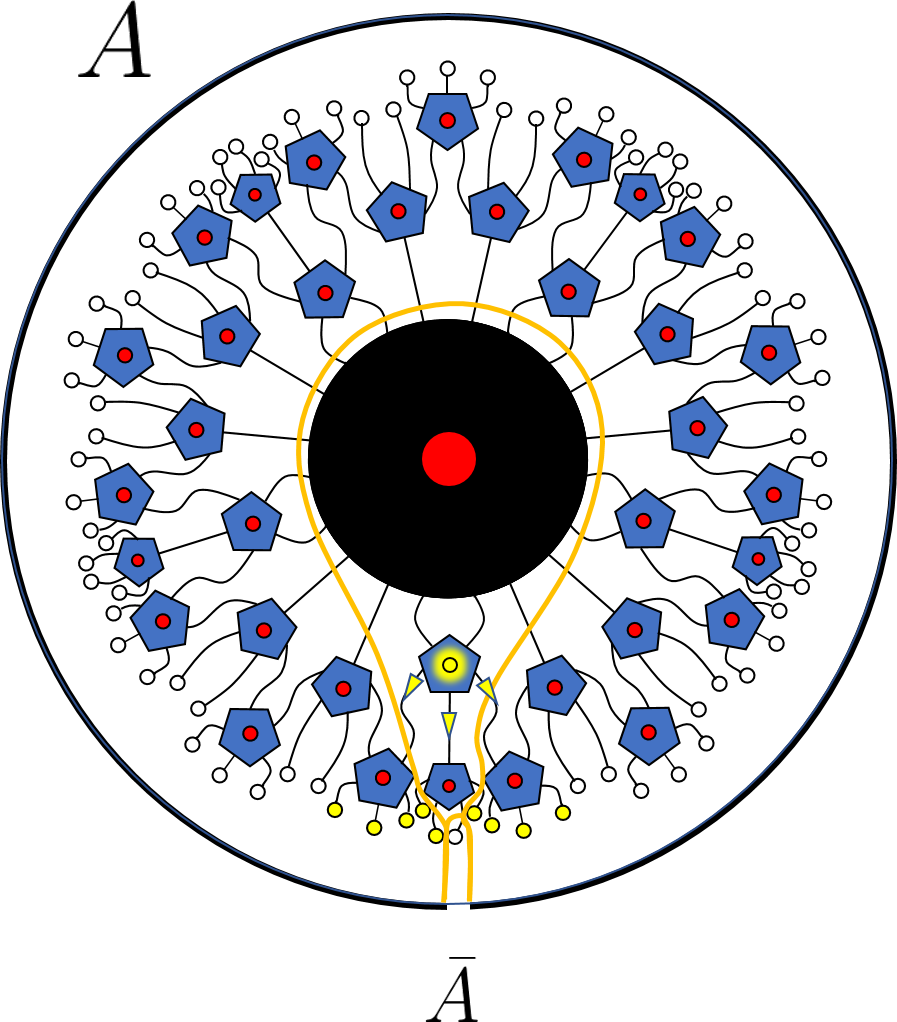}
\centering
\caption{Region $A$ is now sufficiently large that the same operator from Figure \ref{fig:alpha-network}, which is still in region $a'$, can now be pushed directly to the boundary without passing through the black hole. As a result it can be reconstructed on the boundary without any dependence on the state of the black hole.}
\label{fig:qec-network}
\end{figure}
Since the pentagon code does not always satisfy the Ryu-Takayanagi formula this should not disturb us; however, in this case the explanation is quite simple. Because the bulk dangling legs in the pentagon code also have dimension $d$, we cannot ignore their contribution to the bulk entropy. If region $a'$ has $k$ tensors outside the black hole, these can give an additional contribution of $k \log d$ to the bulk entanglement entropy. This means that the quantum extremal surface can be pushed to exclude $a'$ even if we only have
\begin{align}
\log D > \left(n_{a}+n_a^{\text{ext}}-n_{\bar a}-n_{\bar a}^{\text{ext}}-k\right) \log d.
\end{align}
In the case of Figure \ref{fig:alpha-network}, we have $k=n_{a'}=2$ and so this is sufficient to explain the discrepancy; however, it is possible to choose the region $A$ of the boundary such that operators in region $a'$ can only be decoded for smaller $\alpha$ than we expect from AdS/CFT (even taking the bulk entanglement fully into account). Just as with ordinary quantum error correction, the pentagon code gives a weaker $\alpha$-bit encoding than true AdS/CFT.

As in AdS/CFT, if we make the region $A$ sufficiently large, we are able to simulate the bulk for an arbitrarily large space of microstates, or even for a maximally-entangled black hole. As shown in Figure \ref{fig:qec-network}, eventually, as $A$ gets larger, it becomes possible to push operators in region $a'$ to the boundary without having to push them through the black hole. This gives a boundary operator that works exactly for any black hole state (or even no black hole at all).

\section{The space of black hole microstates} \label{sec:microspace}
Throughout this paper and particularly (and explicitly) in Section \ref{sec:entwedge}, we have  assumed that there exists a linear subspace
$$\mathcal{H}_{\text{code}} \cong \mathcal{H}_{BH} \otimes \mathcal{H}_{\text{ext}}$$
 with dimensions $d_{BH}$ and $d_{\text{code}}$ satisfying
\begin{align} \label{eq:limitdim}
\lim_{G_N \to 0} 4 G_N \log d_{BH} = \lim_{G_N \to 0} 4 G_N d_{\text{code}} = \mathcal{A}_0
\end{align}
where we can treat $\mathcal{H}_{BH}$ as describing the microstate of the black hole, while $\mathcal{H}_{\text{ext}}$ describes the state outside the horizon. Is this a reasonable assumption?

In this section we try to argue as rigourously as possible that the answer is yes. Anyone who already agrees with this claim may well find the arguments overly detailed and pedantic. We encourage such readers to skip this section. However, since this paper is built around effects that are only possible when the dimension of the code space is exponentially large, it is important to justify that such exponentially large code spaces do in fact exist.

The AdS/CFT dictionary maps pure black hole microstates in the bulk to pure states that are thermalised on the boundary. We would not necessarily expect the subset of a Hilbert space consisting of approximately thermal pure states to form a linear subspace. There may well be ways to take superpositions of thermal states and produce out-of-equilibrium unthermalised states. As a result, it seems likely that there does not exist a linear subspace of CFT states which contains `all' black hole microstates, without also including some states with very different properties to generic microstates.\footnote{A more precise statement might be that there do not necessarily exist code spaces containing only black hole microstates such that the projection of the thermal ensemble onto the code space approaches the thermal ensemble in the semiclassical limit. It should nonetheless be possible to include all microstates in a sufficiently small energy window without including unthermalised states, as we argue below.}

However attempting to include every black hole microstate in a subspace is much harder than finding a subspace with only black hole microstates and the correct entropy to leading order in $G_N$. If
\begin{align}
\log d_{BH} = \left(1 - \delta\right) \frac{\mathcal{A}_0}{4G_N}
\end{align}
for some fixed very small $\delta > 0$, only a tiny fraction of the microstates, exponentially small in $\frac{1}{G_N}$, need be included. However, restricting to such a subspace only affects $\alpha$ by $O(\delta)$ and can be ignored to leading order.\footnote{Indeed, if we know that code spaces with the correct properties exist in the limit $G_N \to 0$ for any $\delta > 0$, we can construct a sequence of such code spaces where $\delta \to 0$ (albeit potentially very slowly) as $G_N \to 0$. This sequence of code spaces will exactly satisfy \eqref{eq:limitdim}. }
It follows that it doesn't matter whether we are talking about the $\alpha$-bits of this smaller space or the hypothetical much larger space of \emph{all} black hole microstates (if such a space could even be defined, as discussed above).

Let us start by considering the microcanonical ensemble: the maximally mixed state formed from all energy eigenstates within some small energy window whose width is independent of $G_N$ (we of course need the energy itself to scale as $1/G_N$ in order to keep the size of the black hole constant in AdS units). The dimension of this space will obviously scale as $e^{\mathcal{A}_0/4G_N}$.

By the arguments made in \cite{popescu2006entanglement}, almost all states, in the subspace spanned by eigenstates in this energy window, will look almost indistinguishable from the microcanonical ensemble (which in turn will look like the thermal or canonical ensemble to leading order) so long as we only look at the state restricted to a subregion with less than half the size of the entire boundary. In fact, the measure concentration is so strong that we can use the union bound to show that, with very high probability, this will be true for all states in a randomly-chosen subspace of dimension $$e^{(1- \delta) \frac{\mathcal{A}_0}{4 G_N}}$$ in the limit of small $G_N$ for any $\delta > 0$. See, for example, similar arguments in \cite{hayden2012weak}.

This is essentially all we need to construct the space $\mathcal{H}_{BH}$: any such space has the requisite size and all the states are indistinguishable from a thermal state (a black hole) if we look at less than half the boundary, and hence an entanglement wedge that does not include the black hole. It follows that all operators acting on this space approximately commute with any bulk operator in any such entanglement wedge. There is a small entanglement shadow around the black hole horizon which is not contained in any such entanglement wedge, however, it is sufficient for our purposes that the degrees of freedom in $\mathcal{H}_{BH}$ be localised within this shadow, since by definition no entanglement wedge can enter it.  

Furthermore, if we assume the eigenstate thermalisation hypothesis \cite{deutsch1991quantum, srednicki1994chaos, marolf2013gauge}, then the expectation of all simple correlation functions of bulk operators outside the horizon, even ones within the entanglement shadow, will be approximately equal to the thermal expectation for all eigenstates. In fact since the variance of the off-diagonal terms is suppressed by $e^{-S}$, this will still be true for all superpositions of at most $e^{(1-\delta)S}$ eigenstates. An advantage of this construction is that it allows us to explicitly construct the code space simply by restricting to a very small energy window (of size $e^{- \delta S}$). The argument based on measure concentration only allows us to make statements about `most' randomly chosen subspaces.

We can therefore construct a space of sufficient size with no bulk degrees of freedom outside the black hole horizon. Every state in the space is an ``equilibrium state'' \cite{haag1967equilibrium} i.e. they satisfy the KMS condition \cite{kubo1957statistical}. Let us assume we have constructed such a subspace, which we shall call $\mathcal{H}_{BH}^0$. To extend our code space to include $\mathcal{H}_{\text{ext}}$ and hence degrees of freedom outside the black hole requires slightly more work, but essentially it is the same procedure used to construct a code space associated to a single microstate, where we simply add states to the code space if they can be produced by acting with bulk operators on the microstate (up to some limit). 

Suppose we do this for an entire basis of states $\{\ket{\Omega_i}\}$ in $\mathcal{H}_{BH}^0$. We get a set of code subspaces $\{\mathcal{H}_{i,\text{ext}}\}$ -- one for each basis microstate. However, because we already argued that the expectation of any bulk operator is constant within $\mathcal{H}_{BH}^0$, these individual code subspaces will all be (approximately) orthogonal within the larger CFT Hilbert space. We also have the canonical isomorphisms
$$ a_{\text{bulk}} \ket{\Omega_i} \cong a_{\text{bulk}} \ket{\Omega_j} $$
for all bulk operators $a_{\text{bulk}}$. We can use these to identify
\begin{align}
\oplus_i \mathcal{H}_{i,\text{ext}} \cong \mathcal{H}_{BH} \otimes \mathcal{H}_{\text{ext}},
\end{align}
where 
\begin{align}
\mathcal{H}_{BH} \cong \mathcal{H}_{BH}^0
\end{align}
and
\begin{align}
\mathcal{H}_{\text{ext}} \cong \mathcal{H}_{i,\text{ext}}.
\end{align}
Operators on $\mathcal{H}_{\text{ext}}$ act on the code spaces $\mathcal{H}_{i,\text{ext}}$ while operators on $\mathcal{H}_{BH}$ map those spaces into one another in accordance with their action on the basis states $\{\ket{\Omega_i}\}$. 

Any low-energy bulk operator outside the horizon preserves the code spaces $\mathcal{H}_{i,\text{ext}}$, and hence corresponds to an operator acting only on $\mathcal{H}_{\text{ext}}$. Meanwhile operators on $\mathcal{H}_{BH}$ change the state of the black hole itself, while leaving the state outside the horizon unchanged. We have constructed $\mathcal{H}_{\text{code}}$ as desired.

It is important to note that, while many of the steps that we have made in this section are only true approximately, in most cases the corrections are suppressed by powers of $\text{exp}\left(-\delta \mathcal{A}_0 / 4G_N\right)$. As a result, almost all the statements in this section should be perturbatively exact to all orders in $G_N$ for any fixed $\delta > 0$.

The only approximation that is not exact at the level of the perturbative expansion is the equivalence between the microcanonical ensemble of all states in a narrow energy range and the canonical or thermal ensemble. We shall briefly explain this observation, mostly to make it clear that it does not prevent our construction from being perturbatively exact.  If we accept the eigenstate thermalisation hypothesis, any bulk operator $\mathcal{O}$ can be written in the energy eigenbasis as
$$
f_{\mathcal{O}} (E_i) \delta_{ij} + O(e^{-S}).
$$
The expectation of $\mathcal{O}$ for the microcanonical ensemble is just $f_{\mathcal{O}} (E)$. For the canonical ensemble it is given by
\begin{align}
\int dE \,\,\,e^{S(E)- \beta E} f_{\mathcal{O}} (E).
\end{align}
To leading order in a saddle point expansion this agrees with the microcanonical ensemble. However, higher-order corrections will break this equality and are only polynomially suppressed in $1/S \sim G_N$. Fortunately this disagreement is completely unproblematic for our purposes; all we wanted is for every state in $\mathcal{H}_{BH}^0$ to be indistinguishable at the perturbative level outside the horizon -- and for their leading order geometry to be a black hole. Both the canonical and microcanonical ensembles are described by a black hole geometry to leading order in $G_N$, and both will have perturbative corrections and fluctuations in the geometry from the excitation of fields (including gravitons) outside the horizon. The fact the bulk correlators have perturbative differences just tells us that those perturbative bulk corrections are different for the microcanonical and canonical ensembles.

For temperatures below the AdS scale, the black hole saddle point is no longer dominant and so the canonical ensemble is dominated by states consisting of thermal excitations on a vacuum-AdS background, as is the microcanonical ensemble at sufficiently low energies. Nevertheless, there still exists a saddle point in the partition function that becomes semiclassical in the limit $G_N \to 0$. As a result, we expect all the arguments made above to continue to apply, with the caveat that the ensembles being discussed are no longer, strictly speaking, canonical or microcanonical.

\section{Discussion} \label{sec:discuss}
\subsection{Upper bounds on error correction accuracy}
The results in this paper can be used to put strong upper bounds on the accuracy with which error correction can be achieved in certain circumstances. The essential idea is that, if some combination of code space and boundary region forms an $\alpha$-bit code, but not an $(\alpha+\eta )$-bit code, we can place a hard lower bound on the error of the $\alpha$-bit encoding. Since an $\alpha$-bit code can be viewed as a family of ordinary error correcting codes, one for every subspace of sufficiently small dimension, this also puts a lower bound on the error of the ordinary quantum error correcting codes associated with each of the subspaces.

The sudden emergence of previously tiny, non-perturbative errors to create dramatic, leading-order effects is reminiscent of the expected breakdown of Hawking's black hole evaporation calculation at the Page time. Indeed we saw in Section \ref{sec:hawking} that they are two examples of the same phenomenon. No `new' effect causes the change in qualititative behaviour; instead the tiny corrections that have always been present build on top of each other more and more until they suddenly become significant.

We again consider a code space $\mathcal{H}_{\text{code}} \in \mathcal{H}_A \otimes \mathcal{H}_{\bar{A}}$ of states which all have a single black hole of fixed size in the centre of the AdS bulk. We shall use the same notation given in Figure \ref{fig:bhentwedge}. For simplicity we will assume that all states in the code space are identical in region $\bar a$. This avoids the need to deal with the details of subsystem error correction. However, the generalisation of the proof to subsystem or operator algebra error correction is straightforward. Let
\begin{align} \label{eq:sec8.1delta}
\delta = \sup_{\ket{\psi}} \left\lVert \psi_{R\bar{A}} -  \psi_R \otimes \omega_{\bar{A}} \right\rVert_1,
\end{align}
where $\omega$ is the maximally mixed state in the code space be the uncertainty in the forgetfulness for a reference system of fixed dimension $d_R$. We know that $\delta$ grows at most linearly with the dimension $d_R$ of the reference system \cite{hayden2012weak}. 

Let $$\alpha < \frac{\mathcal{A}_2 - \mathcal{A}_1}{\mathcal{A}_0}< \alpha + \eta,$$ for some small $\eta$. Suppose that with $d_R = e^{\alpha S}$, where $S = \mathcal{A}_0 / 4 G_N$ as usual, we have $$\delta < e^{-\eta' S},$$ for some $\eta' > \eta$. Then, for $d_R = e^{(\alpha+\eta) S}$, we would have to have
\begin{align} \label{eq:toosmallerror}
\delta < e^{(\eta - \eta') S} \ll 1.
\end{align}
However this would mean that region $A$ forms an $(\alpha + \eta)$-bit code for region $a'$, which was shown to be impossible in Section \ref{sec:entwedge}. It follows that for $d_R = e^{\alpha S}$ we must have
$$ \delta \geq e^{-\eta' S}.$$
A lower bound on the trace distance can be converted into a lower bound on relative entropy using Pinsker's inequality. There must therefore exist a state $\ket{\psi} \in \mathcal{H}_{\text{code}} \otimes \mathcal{H}_R$ such that
\begin{align}
S\left(\psi^{\bar{A}R} \lVert \omega^{\bar{A}} \otimes \psi^R \right) \geq \frac{1}{2} e^{-2 \eta' S}.
\end{align}
Since the entanglement wedge of $\bar{A}R$ does not include the black hole for either state, the bulk relative entropy is zero. As a result this provides a strict upper bound on the accuracy of the equality between bulk and boundary relative entropies (\ref{eq:relent}), even when the bulk relative entropy is zero. 

Similarly, using (\ref{eq:forgetful}), we see that this provides an upper bound on the accuracy of the error correction for the $\alpha$-bit code of
\begin{align}
\left\lVert \mathcal{D} \circ \Tr_{\bar{A}} - \Id \right\rVert_\diamond \geq \frac{1}{64} e^{-2 \eta' S}.
\end{align}
Of course this argument, as we have given it, only puts a lower bound on the largest recovery error for any state in the space. In the Heisenberg picture, it lower bounds the worst-case error in operator recovery. That operator may be some complicated operator involving the black hole Hilbert space $\mathcal{H}_{BH}$. 

However, even if we restrict ourselves to the algebra $\mathcal{A}$ of bulk operators in region $a'$, but outside the black hole horizon, we can make an almost identical argument to the one above, by using Theorem \ref{thrm:alinfdist} from Appendix \ref{sec:algebra}. The uncertainty $\delta_{\mathcal{A}}$ (defined formally in \eqref{eq:uncertaintyforgetalgebra}) with which the algebra $\mathcal{A}$ is forgotten by region $\bar A$  still grows at most linearly with the allowed dimension of the reference system. Hence, if the algebra could be error corrected too accurately for all code subspaces of dimension $e^{\alpha S}$ (which would imply that $\delta_{\mathcal{A}}$ is similarly small, even for larger code spaces, so long as we restrict the reference dimension to be at most $e^{\alpha S}$), it could also be corrected for all larger spaces (which we know to be impossible) with the same bounds as above. The same argument can be made even for the algebra generated by any single operator. It follows that we get the same lower bound on the error for every operator in region $a'$, whether acting on the black hole or only acting in the bulk outside the horizon.

By either tuning the region $A$ or adjusting the value of $\alpha$, we can make the parameter $\eta$ (and hence $\eta'$) arbitrarily small. It follows that there exist cases where error correction is possible if $G_N \to 0$, but that at large but finite $G_N$ the smallest achievable error is greater than $\text{exp}(-\eta/G_N)$ for any arbitrarily small $\eta > 0$.\footnote{Technically, the error should presumably smoothly interpolate between exponentially suppressed and order one as the dimension is increased. However, we do not have semiclassical control over the effects of such fine-grained changes in code space dimension.} Similarly there exist states whose bulk relative entropy is zero, but whose boundary relative entropy is at least $\text{exp}(-\eta/G_N)$ for any $\eta > 0$.

The equality between bulk and boundary relative entropies is not true to all orders in $G_N$; however, a more precise equality was developed in  \cite{dong2018entropy} that is true to all orders in perturbation theory. If the bulk relative entropy is exactly zero, this reduces to the statement that the boundary relative entropy should be perturbatively equal to zero to all orders in $G_N$. Even this more sophisticated equality was always expected to be broken by non-perturbative effects. However, here we have shown that such effects are absolutely required for the expected error-correction properties of AdS/CFT to hold -- at least when a black hole is involved.

If there is no black hole, our arguments do not apply. However, a simple argument \cite{kelly2017bulk} based on the Reeh-Schlieder theorem \cite{reeh1961bemerkungen} shows that it is still impossible for the error correction to be exact, even for a code space based on perturbations of the vacuum state $\ket{\Omega}$. Divide the boundary into four non-empty disjoint regions $A$, $B$, $C$ and such that some bulk operator $\phi$ lies in the entanglement wedge of $AB$ and the entanglement wedge of $BC$, but not in the entanglement wedge of $B$. There exists a boundary operators $\phi_{AB}$ and $\phi_{BC}$ acting on $AB$ and $BC$ respectively such that
\begin{align} \label{eq:roughly}
\phi_{AB} \ket{\Omega} \simeq \phi \ket{\Omega} \simeq \phi_{BC} \ket{\Omega}.
\end{align}
Since $\phi$ does not lie in the entanglement wedge of $B$ we know that $\phi_{AB} - \phi_{BC} \neq 0$. However, since $D$ is non-empty, then, by the Reeh-Schlieder theorem, we know that
\begin{align}
(\phi_{AB} - \phi_{BC}) \ket{\Omega} \neq 0
\end{align}
and hence the equalities in \eqref{eq:roughly} cannot be exact. We cannot ascribe the approximate nature of the error correction in AdS/CFT purely to the presence of particular black hole microstates. However it is only in the presence of a black hole that we are able to place lower bounds on the size of the error.

What causes these non-perturbative corrections? Clearly, it seems to have something to do with the existence of a closely competing candidate for the entanglement wedge. The exponents in the error bounds that we calculate are proportional to the difference between the Ryu-Takayanagi formula, evaluated for each of the two candidate wedges.  This has the strong feel of a contribution from a subleading saddle point calculation. Unfortunately, it is not clear how this such contributions should be calculated, at least in the usual approach to deriving the RT formula \cite{lewkowycz2013generalized, faulkner2013quantum}. Such effects are, nonetheless, widely expected to exist (for example, in order to smooth out phase transitions in the entanglement entropy at finite $N$).

\subsection{State dependence of the entanglement wedge}
If we consider a code space $\mathcal{H}_{\text{code}}$ of perturbations about vacuum-AdS, the entanglement wedge of a region $A$ of the boundary is state-independent in the semiclassical limit where gravity decouples and the background geometry is fixed. We can of course consider superpositions of states with different background classical geometries. However, since gravity decouples, these become non-interacting superselection sectors of the theory. The entanglement wedge is fixed for each classical geometry and hence does not depend on the state in any non-trivial sense.

In contrast, at finite coupling, the geometry is dynamical (and the bulk entanglement term in the RT formula may in principle compete with the area term). As a result, the entanglement wedge will in general depend on the state -- the size of the fluctuations will be approximately Planckian scale. We will be unable to reconstruct operators precisely unless they are separated from the entanglement wedge by a large distance in Planck units.

Alpha-bit codes for a black hole in the limit $G_N \to 0$ are in some sense an intermediate case: the geometry is fixed and smooth at arbitrarily small scales. However, large amounts of bulk entanglement can cause the entanglement wedge to be state-dependent. If we believe in the ER=EPR correspondence \cite{maldacena2013cool}, this bulk entanglement may itself have a geometrical interpretation. 

This provides arguably the most controlled setting in which the entanglement wedge is state-dependent. We do not have to deal with any issues of backreaction or fluctuations in the geometry. By making the coupling $G_N$ small, we can make the size of the $\alpha$-bit region (which was shown in Section \ref{sec:btz} to be approximately the AdS scale) arbitrarily large compared to the Planck scale where the classical geometry breaks down. In this way we are able to isolate the issues that arise from state dependence of the entanglement wedge, without having to deal with all the attendant issues of the classical geometry breaking down at the Planck scale.

Now consider the quantum error correction which exists in each of these three cases. In the strict $G_N \to 0$ limit and the absence of black holes, the quantum error correction approaches zero error. Conversely, at finite coupling, the error correction will always only be approximate. Just like when we looked at the fluctuations of the engtanglement wedge, the $\alpha$-bit encodings discussed in this paper form an interesting intermediate case. The actual error $\varepsilon$ tends to zero for any decoding of allowed subspaces in the semiclassical limit. However, the phenomenon is only possible at all because the error is non-zero for any finite dimensional code space and finite coupling -- it is still a phenomenon inherent only to approximate quantum error correction. 

It seems likely that, if we want to truly understand entanglement wedge reconstruction in the context of a dynamical spacetime geometry, we may well have no choice but to think hard about issues specific to approximate quantum error correction. 

\subsection{State dependence of operators and the Papadodimas-Raju proposal} \label{sec:statedepend}
Of course, in $\alpha$-bit codes, it is not only the entanglement wedge that is state-dependent; operators within the `$\alpha$-bit region' (region $a'$ in Figure \ref{fig:bhrentwedge}) are as well. If we wish to construct a decoding map $\mathcal{D}: S(A) \to S(a \otimes a')$ such that
\begin{align}
\left\lVert\mathcal{D} \circ \Tr_{\bar{A}}(\cdot) - \Tr_{\bar{a}}(\cdot) \right\rVert_\diamond
\end{align}
is small and hence we can recover the state on $a \otimes a'$ from the state on $A$, we must first choose some code subspace of dimension less than $d^\alpha$ that we wish to decode -- it is impossible to construct an approximate decoding channel that works for all states at once.
In the Heisenberg picture, an operator $\phi_{a'}$ that is local to region $a'$ can be approximately simulated by an operator $\mathcal{D}^\dagger ( \phi_{a'})$ on the boundary that is local to $A$ \textit{so long as} we only require that this operator behave correctly when restricted to some code subspace of dimension less than $d^\alpha$.

This state dependence of operators is unavoidable: there is no operator that will work for all states at once. This is in contrast to previous work on entanglement wedge reconstruction where there was apparent state dependence because the method of constructing the operator made use of a choice of fixed state, but where this was merely an artifact of the construction. Boundary operators would still work for the entire code space, even if the boundary operator corresponding to a given bulk operator was not unique and so could have a definition that depended non-trivially on a choice of state.

One open question is to what extent similar state dependence exists for operators reconstructed on boundary subregions, even in the absence of a black hole, but at finite coupling. In this case fluctuations in the entanglement-wedge are only of order the Planck scale and as a result it is not even clear how well-defined it is to talk about bulk operators localised \textit{within} such a region. Nonetheless, brushing aside such issues, one might speculate that it should again prove possible to reconstruct operators near the RT surface with greater accuracy if they only have to act correctly on a single state rather than an entire large code space of states. We will not, however, do so any more here.

More importantly, it has been argued, primarily by Papadodimas and Raju \cite{papadodimas2014state, papadodimas2013infalling}, that local operators behind a black hole horizon are necessarily state-dependent. For some fixed choice of KMS-equilibrium black hole state $\ket{\psi_0}$ (essentially a pure state in the space $\mathcal{H}_{BH}^0$ that we defined in Section \ref{sec:microspace}), they use a variant of Tomita-Takesaki theory to construct `mirror operators' that depend on $\ket{\psi_0}$ and that they claim describe the modes behind the black hole horizon. The mirror operators will behave correctly, so long as the state lies in the `small Hilbert space' formed by applying simple operators to the chosen equilibrium black hole state $\ket{\psi_0}$. 

These small Hilbert spaces form the equivalent of the decodable code subspaces in $\alpha$-bit codes. Since their dimension is independent of $G_N$, in our language it is tempting to say that the Papadodimas-Raju proposal effectively gives a procedure for recovering the zero-bits of the interior of the black hole.\footnote{One apparent difference between the Papadodimas-Raju proposal and zero-bits is that the small Hilbert space is not a completely arbitrary choice of subspace. Instead we are free to choose to choose the equilibrium state $\ket{\psi_0}$, and this choice determines the entire small Hilbert space. However, if we want to reconstruct simple interior operators then we clearly need those simple operators to preserve the code subspace. So this constraint on the small Hilbert space is simply a consequence of the fact that in $\alpha$-bit codes it is only meaningful to reconstruct operators for a subspace that is preserved by those operators.}

There has been considerable debate as to whether the Papadodimas-Raju proposal is consistent with a standard quantum mechanical interpretation of the bulk e.g. \cite{harlow2014aspects}. In our construction there are no such concerns: the state dependence is simply a phenomenon that arises from this particular form of quantum error correction and because we are restricting ourselves to a subregion of the boundary. If we have access to the entire boundary, the entire bulk (outside the horizon) can be understood without resorting to any state dependence beyond the basic geometry of the bulk itself. In contrast, in the Papdodimas-Raju proposal, the observer always has access to the entire boundary; in this sense there aren't any errors to correct. If we accept the usual paradigm for error correction in AdS/CFT, where there is an isometry from a bulk code space to a larger boundary Hilbert space, then all operators, including interior operators should be reconstructable on the entire boundary in a state-independent way.

However there are important physical reasons, independent of Papadodimas-Raju construction itself, to think that bulk reconstruction is only possible for interior modes in a state-dependent way. Recent work \cite{kourkoulou2017pure} has shown that in a simple toy model of quantum gravity, known as the SYK model, there exists an (over)complete basis of black hole microstates for which simple excitations behind the horizon can be pulled out of the horizon by using a perturbed Hamiltonian which depends on the microstate in question. It follows that we can decode interior operators by evolving exterior operators with this perturbed Hamiltonian. However the reconstructed interior operator that one obtains will depend on the black hole microstate. Even more recently, it was argued in \cite{de2018interior} that it should be possible in a general holographic CFT to use the mirror operators $\tilde O$ to create a negative energy shockwave that will pull degrees of freedom outside the horizon -- provided the state lies in the small Hilbert space defined above. 

On the other hand, there clearly cannot be a single Hamiltonian that can pull out interior modes for \emph{any} black hole microstate. If such a Hamiltonian existed, we could presumably use it to evolve one CFT in a thermofield double state and thereby create a traversable wormhole without having any interaction between the two CFTs. From this, it is very natural to conclude that the modes behind the horizon can only be defined on the boundary in a state-dependent way. The Papadodimas-Raju proposal suggests that only code subspaces of fixed dimension can be decoded -- that perhaps \emph{only} the zero-bits of the black hole interior are encoded in the boundary CFT.

It is tempting to go further and suggest that these two examples of state dependence -- the $\alpha$-bit codes discussed in this paper and the possible state dependence of operators in the black hole interior -- result from the same basic mechanism. In the situations considered in this paper, the region $a'$ where operators are state-dependent lies behind the causal (Rindler) horizon, just as the state dependence in the Papadodimas-Raju proposal appears for operators behind the black hole horizon.  Furthermore, if we consider too large a code space of interior bulk states, it seems plausible that the RT surface might jump to near the horizon, for states sufficiently entangled with a reference system. Hence it would follow that interior modes could only be reconstructed for sufficiently small code subspaces, thus `deriving' something like the Papadodimas-Raju proposal from the same sort of argument we have been making throughout this paper.

However, if we really take this argument seriously, we don't reach quite the same conclusion as Papadodimas and Raju. For a single-sided pure black hole the true Ryu-Takayanagi surface for the entire boundary should trivially be empty since the boundary state is pure. Even if we have a significant amount of bulk entanglement between the region behind the horizon and a reference system, the entropy of the complete boundary CFT will simply be equal to the entropy of the reference system. In order to make the Ryu-Takayanagi surface lie directly on the black hole horizon we would need a reference system of dimension 
$$d_R\sim e^{S}.$$ 
Hence bulk reconstruction for code spaces whose dimension 
$$d_S\sim e^{\alpha S},$$
for any $\alpha < 1$, which would be far larger than the $G_N$-independent small Hilbert space in the Papadodimas-Raju proposal. 

One speculative way to try to reconcile these differences would be to argue that we can include the mirror operators for an increasing number of modes (perhaps because of the reduced backreaction) as we take $G_N \to 0$, creating a Hilbert space whose dimension scales rapidly with $1/G_N$. An alternative approach, which is perhaps more compelling, would be to conclude that the Papadodimas-Raju proposal is correct but incomplete -- that there also exist more complicated reconstructions which work for larger code subspaces. In particular it should be possible to combine the Papadodimas-Raju mirror operators $\tilde{O}
_S$ for a large number of orthogonal small Hilbert spaces $\mathcal{H}_S$ with the boundary projectors $P_S$ associated to each small Hilbert space to construct an operator
$$\tilde{O} =  \sum_S \tilde{O}_S P_S,$$
that reproduces the bulk interior mode operator for a far larger class of states. This argument will break down when the small Hilbert spaces stop all being approximately orthogonal, but (by similar arguments to those in Section \ref{sec:microspace}) we might hope that this should not occur if we only consider $e^{(1-\delta) S}$ small Hilbert spaces for any fixed $\delta > 0$. (Note that even if you can decode operators directly behind the black hole for any code subspace of dimension $e^{(1-\delta) S}$ for arbitrary small $\delta > 0$, it is still not possible to decode operators directly behind the black hole for the thermofield double state, which, as we argued above, should clearly be impossible.\footnote{The idea that the $\alpha$-bits are encoded for any $\alpha < 1$, but not for $\alpha = 1$, is strongly reminiscent of the discontinuity in the amortised $\alpha$-bit capacity at $\alpha = 1$, see \cite{alphabits}.})

Similarly, the method used in \cite{kourkoulou2017pure} to pull operators out from behind the horizon in the SYK model relies on evolving the system with a perturbed Hamiltonian precisely tuned to the black hole microstate in question. However, even if the Hamiltonian were only approximately tuned to the microstate (for example if only some sufficiently large fraction of the terms in the perturbed Hamiltonian were tuned correctly), it should still be possible to successfully pull the mode from behind the horizon. Hence a single Hamiltonian, and thus a single boundary reconstruction, can be used for a larger subspace of microstates. It would be very interesting to try to calculate exactly how large a subspace could successfully be reconstructed in this way.

As our understanding of AdS/CFT has developed, there has been a gradual de-emphasis of the idea of a unitary isomorphism between two distinct bulk and boundary Hilbert spaces, as in the traditional sense of a duality. Since most boundary states correspond to a large black hole with little geometrical bulk interpretation, it is instead more natural to think of an isometry from a code space of semiclassical geometrical bulk states to the larger boundary space.\footnote{It is of course still possible that there exists a non-perturbative bulk description of the entire Hilbert space. In this case we would have a true duality between different theories and the code space isometries would simply come from the embedding of semiclassical sectors into the larger bulk Hilbert space. As yet, however, such a complete non-perturbative bulk description remains unknown and it is not clear that one must exist at all.} By recognising that only a (relatively) small code space of states have any given bulk geometry, apparent paradoxes where bulk operators appeared to commute with every local boundary operator were resolved \cite{almheiri2015bulk}. However, if operators behind a black hole horizon are truly necessarily state-dependent -- despite the fact that we have access to the entire boundary -- then we cannot even have a simple isometry from bulk to boundary, since an isometry can always be error-corrected exactly without any need for state dependence. 

One possibility is that we need to consider a general quantum channel, where bulk operators outside the horizon can lie in the image of the space of boundary operators under the adjoint channel, but bulk operators inside the horizon can only be simulated in a state-dependent way. This more general formalism was used in \cite{cotler2017entanglement}, although any physical implications of a noisy bulk-to-boundary channel were not discussed and the possiblility was included mostly for the sake of completeness. However a noisy quantum channel does not seem to be quite the right object. It would suggest that pure bulk states should be associated to mixed boundary states. Instead almost the opposite seems to be the case, with multiple bulk descriptions corresponding to the same pure boundary state. The bulk-to-boundary map is more reminiscent of a linear map that is not always exactly isometric. Certainly, such a non-isometric linear map could appear very naturally in a tensor network model of AdS/CFT. For obvious physical reasons, generalisations of quantum error correction to such maps have not really been studied; they seem to be at least worth considering, if we want to understand holography.

There is one final connection between the Papadodimas-Raju proposal and more general zero-bit codes, which is too intriguing not to mention, but whose exact meaning and significance is somewhat unclear to the authors of this paper. The mirror operators in the Papadodimas-Raju proposal give an effective doubling of the CFT degrees of freedom, provided you only consider sufficiently simple operators. This doubling appears to correspond extremely elegantly with the fact that it is possible to encode zero-bits at an asymptotic rate of at most two zero-bits per qubit. In other words, if we only want the boundary state to encode the zero-bits of the bulk state, we are able to encode twice the number of degrees of freedom into the boundary state -- potentially the modes both in front of and behind the horizon. Of course, as discussed above, the map from bulk to boundary does not quite seem to be a noisy quantum channel in the usual sense. Hence we cannot really think of mirror operators as giving a capacity-achieving zero-bit code. However the correspondence seems too tantalising to completely ignore.

\subsection{Explicit reconstruction of state-dependent bulk operators}
Until this point we have contented ourselves with showing the existence of decoding maps with the desired properties, we have not worried about \emph{how} they should be constructed. Fortunately to a large degree this work has already been done for us. For any particular choice of subspace that we wish to decode, finding an explicit decoding map for an $\alpha$-bit entanglement wedge is no different than the task of finding an explicit decoding map for ordinary entanglement wedge error correction, when the code space is given by the subspace we wish to decode. This task has been achieved recently in \cite{cotler2017entanglement,faulkner2017bulk}.

In \cite{cotler2017entanglement}, it was shown that operators in the entanglement wedge can be reconstructed on the boundary using a universal decoding map, built from the twirled Petz map $\mathcal{R}_{\sigma,\mathcal{N}}$, developed in \cite{junge2015universal}. It is defined for some fixed state $\sigma$, which we shall take to be the maximally mixed state on the subspace we wish to decode, and encoding channel $\mathcal{N}$, which here is simply the partial trace over $\bar A$, by
\begin{align}
\mathcal{R}_{\sigma,\mathcal{N}} = \int_{\mathbbm{R}} dt \, \beta_0 (t) \sigma^{\frac{1}{2}(1 - i t)} \mathcal{N}^\dagger \left( \mathcal{N}(\sigma)^{-\frac{1}{2}(1 - i t)} ( \cdot )  \mathcal{N}(\sigma)^{-\frac{1}{2}(1 + i t)}\right)  \sigma^{\frac{1}{2}(1 + i t)},
\end{align}
where $\beta_0 (t) := \frac{\pi}{2} (\text{cosh}(\pi t) + 1)^{-1}$. The argument in \cite{cotler2017entanglement}, like the arguments made here and in \cite{dong2016reconstruction}, relies on the approximate equality between bulk and boundary relative entropies. However, in this case, the relative entropies under consideration are associated to the boundary region $A$, used to decode the state, rather than its complement $\bar A$. For this argument to go through, we need the bulk region we wish to reconstruct to be contained in the entanglement wedge of $A$ for all states (pure or mixed) in the space we wish to decode. This is exactly equivalent, as we would hope, to the bulk region not being contained in the entanglement wedge of $\bar{A} R$ for any pure state that may be entangled with the reference system $R$. It follows that all the same arguments used in Section \ref{sec:entwedge} still apply and the twirled Petz map can be used to decode correctable subspaces in the $\alpha$-bit codes. In practice unfortunately the twirled Petz map is hard to evaluate even for simple bulk geometries -- trying to do so for a particular subspace of black hole microstates is likely to be essentially impossible. Furthermore, since the argument in \cite{cotler2017entanglement} is based on the equality of bulk and boundary relative entropies for bulk states that are not identical either within or outside the entanglement wedge of $A$, it is unclear whether this recovery map is exact to all orders in perturbation theory.

In \cite{faulkner2017bulk} a different explicit formula for entanglement wedge reconstruction was developed by relating the bulk and boundary modular flows (the evolution of operators using the modular Hamiltonian $K = - \log \rho$). The basic extension of the extrapolate dictionary to bulk and boundary modular evolutions of operators developed in \cite{faulkner2017bulk} should continue to apply when the bulk contains a black hole. However, because the bulk no longer consists only of free fields to leading order (the reconstruction needs to also somehow depend on the state of the black hole), the bulk modular-evolved fields will no longer be simply a Bogoliubov transformation of the original fields. This was the key property used in \cite{faulkner2017bulk} to argue that they formed a natural basis to attempt entanglement wedge reconstruction.

It should be essentially unsurprising that explicit construction of the state dependent operators is highly challenging in practice. The whole point of the requirement of state dependence is that we are unable to reconstruct the operators without taking advantage of our detailed knowledge of the subspace of black hole states in question. In Section \ref{sec:tensor} we saw that in tensor network toy models this involves literally pushing the operator through the black hole; all the details of the construction of the boundary operator depend on the black hole states in question. Nonetheless it is comforting that there exist an explicit, even if impractical, reconstruction procedure.

\subsection{Black holes are $\alpha$-bit sup?} \label{sec:abitsup}
We argued in Section \ref{sec:hawking} that the $\alpha$-bits of a black hole are encoded in any fraction greater than $$p =\frac{1+\alpha}{2}$$ of its Hawking radiation so long as the evaporation is approximately thermodynamically reversible. The $\alpha$-bit capacity of the noiseless qubit channel \cite{alphabits} is equal to  $$\frac{2}{1+\alpha}.$$ It follows that Hawking radiation is a \textit{capacity-achieving} $\alpha$-bit encoding of the black hole state. 

Although less obvious at first glance, this continues to be true in the `entanglement-assisted' case, when some Hawking radiation (with entropy $\beta S$) has been emitted before the unknown black hole is dropped in: in this case we require a fraction greater than 
\begin{align} \label{eq:enthawkingrad}
p = \frac{1 + \alpha - \beta}{2}
\end{align}
of the Hawking radiation in order to decode a subspace of dimension $\text{exp}(\alpha S)$. The `entanglement-assistance' has reduced the fraction of the Hawking radiation that is required to decode a subspace of the same size, so this might seem to exceed the $\alpha$-bit capacity. However, Bob had to know the state of the original black hole, which must have had entropy of at least $\beta S$ in order to be so entangled with the original Hawking radiation. Even if the remaining `unknown' entropy of the black hole were as large as possible, it would only be equal to $(1-\beta) S$. As a result, Bob is effectively only recovering the $\alpha'$-bits of this smaller unknown system for
\begin{align}
\alpha' = \frac{\alpha}{1-\beta}.
\end{align}
He can decode the $\alpha'$-bits of a system with entropy $(1-\beta) S$ using Hawking radiation with entropy only
\begin{align}
p \,S = \frac{1+\alpha'}{2} (1 - \beta) S.
\end{align}
In other words, we find that the $\alpha'$-bits of the unknown black hole are still being encoded at the noiseless $\alpha'$-bit capacity rate of $2/(1+\alpha')$.

If the black hole evaporation is significantly thermodynamically irreversible, it no longer saturates the $\alpha$-bit capacities. However, this inefficiency is an inevitable consequence of thermodynamic entropy being produced. As we saw in \eqref{eq:irreversibleabitcondition}, the black hole still reveals its $\alpha$-bits in the Hawking radiation as soon as the Hawking radiation contains sufficient entropy to both purify the black hole and encode the $\alpha$-bits of the initial state

Similarly, in Section \ref{sec:btz}, we found that the $\alpha$-bits of a large (in AdS units) BTZ black hole are encoded holographically in any fraction of the boundary greater than
$$\frac{1+\alpha}{2}.$$
This is in fact also true in higher dimensions, since we can approximate minimal surfaces around a large black hole by the minimal surface connecting the horizon to the boundary, together with part of the horizon itself. Corrections to this approximation will only be of order the AdS scale and so are negligible in the limit of large horizon area. As a result they have no effect on $\alpha$ to leading order in $L/R$.

If we impose a UV cut-off on the CFT, which cuts off the bulk close to the horizon area, we should expect the effective dimension of the CFT to be only of order $e^S$. As a result, the boundary subregion forms a capacity-achieving encoding of the black hole state. 

Because black holes seem to achieve the $\alpha$-bit capacity whenever possible, we say that black holes are `$\alpha$-bit sup'. As discussed in Section \ref{sec:hawking}, the subspace decoupling duality shows that anything which looks  `as thermal as possible' when looking at less than half of its degrees of freedom is necessarily also $\alpha$-bit sup when one has access to more than half of the degrees of freedom. It should therefore not surprise us at all that black holes tend to satisfy it. However, to a physicist used to thinking of black holes as hiding information whenever possible, it may seem counterintuitive that we can equivalently think of them as revealing information (specifically their $\alpha$-bits) as soon as possible.

At this point, there exist almost no known explicit constructions of capacity-achieving $\alpha$-bit codes;\footnote{The only known example is for the zero-bit capacity of the noiseless cbit channel \cite{fawzi2013low}.} it has only been possible to show that they exist by considering randomly chosen unitaries. It is questionable exactly how much more explicit it is to define a space as `a subspace of black hole states in AdS/CFT' than to define it as `a subspace that is k-forgetful to the environment (which provably must exist)', but it is amusing to note from a quantum information perspective that black holes therefore form the first example of an explicit capacity-achieving $\alpha$-bit code for a noiseless quantum channel (for a forgiving definition of explicit).

\section{Acknowledgements}
We would like to thank Ahmed Almheiri, Daniel Harlow, Matthew Headrick, Isaac Kim, Aitor Lewkowycz, Sepehr Nezami, Phil Saad, Grant Salton, Steve Shenker, Lenny Susskind, Aron Wall, Michael Walter and Edward Witten for valuable discussions. In particular, we would like to thank Jonathan Sorce for discussions on the contents of the acknowledgements section. This work was supported by AFOSR (FA9550-16-1- 0082), CIFAR and the Simons Foundation.

\appendix

\section{Operator algebra quantum error correction and the information disturbance tradeoff} \label{sec:algebra}
The framework of operator algebra quantum error correction was first introduced in \cite{beny2007generalization}; it is a generalisation of the notion of subsystem quantum error correction that is natural in holography because it allows one to talk about superpositions of different geometries in AdS/CFT \cite{almheiri2015bulk, beny2018approximate}. Even though we make essentially no use of the formalism in the rest of the paper, we include a brief review here, mainly focussed on providing a statement of (and also something of the context behind) Theorem \ref{thrm:alinfdist}, which was first proved in \cite{beny2009conditions}.\footnote{For earlier related results, see \cite{tyson2010two}.}

Theorem \ref{thrm:alinfdist}, applied to the special case of a subsystem error correcting code as \eqref{eq:deltaepsilon}, lies at the heart of the results of Section \ref{sec:entwedge}.  We include the operator algebraic version here because it has not appeared previously in the literature on AdS/CFT and is the natural generalisation of the Dong-Harlow-Wall condition \cite{dong2016reconstruction} to approximate reconstruction. Using the argument from \cite{dong2018entropy} that the boundary relative entropy is zero to all orders in perturbation theory if the bulk states agree exactly, Theorem \ref{thrm:alinfdist} is sufficient to prove that entanglement wedge reconstruction can be made exact to all orders in $1/N$. A version of the proof assuming the existence of a tensor product factorization was given in Section \ref{sec:approxent}. We give a more general argument from an algebraic perspective here, although we still need to assume that all the Hilbert spaces involved are finite-dimensional.  

So long as the dimension of the code space grows at most polynomially with $N$ or $1/G_N$, there is no need to consider a reference system and the argument can proceed almost identically to the argument based on assuming exact error correction given in \cite{dong2016reconstruction}. However, as we have seen in this paper, the argument completely fails for larger code spaces unless we additionally consider states entangled with a reference system. Explicitly considering the approximate case therefore provides important non-trivial insight.

A quantum channel $\mathcal{N}:S(\mathcal{H}_1) \to S(\mathcal{H}_2)$ is a completely positive trace-preserving linear superoperator and hence maps density matrices to density matrices. The adjoint channel $\mathcal{N}^\dagger: \mathcal{L}(\mathcal{H}_2) \to \mathcal{L}(\mathcal{H}_1)$ is defined by
\begin{align}
\Tr \left( \mathcal{N}^\dagger ( X ) \rho \right) = \Tr \left( X \mathcal{N}(\rho)\right)
\end{align}
and hence is a unital, trace-preserving map from operators on $\mathcal{H}_2$ to operators on $\mathcal{H}_1$. When working in the Heisenberg picture, this is sometimes given as the definition of a quantum channel.

The diamond norm  between two channels  $\mathcal{N}_1,\, \mathcal{N}_2$ is defined by
\begin{align}
\left\lVert \mathcal{N}_1 - \mathcal{N}_2 \right\rVert_\diamond = \sup_{\lVert\rho\rVert_1 \leq 1} \left\lVert \left[\left(\mathcal{N}_1 - \mathcal{N}_2 \right) \otimes \Id_R \right] \rho \right\rVert_1,
\end{align}
where we can take the dimension of the reference Hilbert space $\mathcal{H}_R$ to be equal to the dimension of $\mathcal{H}_1$. Equivalently
\begin{align}
\left\lVert \mathcal{N}_1 - \mathcal{N}_2 \right\rVert_\diamond = \sup_{X \neq 0} \frac{\left\lVert \left[\left(\mathcal{N}^\dagger_1 - \mathcal{N}^\dagger_2 \right) \otimes \Id_R \right] X \right\rVert_\infty}{\left\lVert X \right\rVert_\infty},
\end{align}
where the supremum is taken over non-zero operators $X$ acting on $\mathcal{H}_2 \otimes \mathcal{H}_R$.

A finite-dimensional von Neumann algebra $\mathcal{A} \subseteq \mathcal{L}(\mathcal{H}_1)$ is a subset of linear operators that contains the identity and is closed under multiplication, scalar multiplication, addition and Hermitian conjugation. It can be shown that for such an algebra, there exists an orthogonal decomposition of the Hilbert space $$\mathcal{H}_1 = \bigoplus_i \mathcal{H}_{A_i} \otimes \mathcal{H}_{\bar A_i}$$ such that the algebra $\mathcal{A}$ consists of all operators of the form
$$
X = \sum_i X_i \otimes \mathbbm{1}_{\bar A_i}
$$
for $X_i \in \mathcal{L}\left(\mathcal{H}_{A_i}\right)$. If $P_i$ is the projector onto $\mathcal{H}_{A_i} \otimes \mathcal{H}_{\bar A_i}$, we define the projector $\mathcal{P}_{\mathcal{A}}$ onto the algebra to be the quantum channel
\begin{align}
\mathcal{P}_{\mathcal{A}} (\rho) = \sum_i \frac{1}{d_{\bar A_i}} \Tr_{\bar A_i} \left( P_i \,\rho\, P_i \right) \otimes \mathbbm{1}_{\bar A_i},
\end{align}
Note that $\mathcal{P}_{\mathcal{A}}^2 = \mathcal{P}_{\mathcal{A}} = \mathcal{P}_{\mathcal{A}}^\dagger$. $\mathcal{P}_{\mathcal{A}}$ is therefore both trace-preserving and unital; it is the quantum channel in both the Schr\"{o}dinger and Heisenberg pictures. For notational convenience we write $\rho_{\mathcal{A}} = \mathcal{P}_{\mathcal{A}} (\rho)$. The image of the channel $\mathcal{P}_{\mathcal{A}}$ (when acting on density matrices) is the intersection of the algebra $\mathcal{A}$ with space of density matrices $S(\mathcal{H})$ and is canonically isomorphic to the space of positive normalised linear functionals on $\mathcal{A}$, which is the standard abstract definition of states on a von Neumann algebra.

We say that the pair of channels $\mathcal{N}: S(\mathcal{H}_1) \to S(\mathcal{H}_2)$ and $\mathcal{D}: S(\mathcal{H}_2) \to S(\mathcal{H}_1)$ form an exact quantum error correcting code for the finite-dimensional von Neumann algebra $\mathcal{A}$ if 
\begin{align}
\mathcal{D} \circ \mathcal{N} = \mathcal{P}_{\mathcal{A}}.
\end{align}
Note this implies that for all $X \in \mathcal{A}$ then
\begin{align}
\left(\mathcal{D} \circ \mathcal{N}\right)^\dagger X = \mathcal{P}_{\mathcal{A}}(X) = X.
\end{align}
To make this approximate, we simply allow some small separation $\varepsilon_{\mathcal A}$ in terms of the diamond norm
\begin{align}
\left\lVert \mathcal{D} \circ \mathcal{N} - \mathcal{P}_{\mathcal{A}} \right \rVert_\diamond \leq \varepsilon_{\mathcal A}.
\end{align}

Define the von Neumann algebra $\mathcal{A}'$ to be the set of operators that commute with all operators in $\mathcal{A}$ (this is known as the commutant of $\mathcal{A}$). These are the operators of the form 
\begin{align}
X' = \sum_i \mathbbm{1}_{A_i} \otimes X'_i
\end{align}
for $X'_i \in \mathcal{L}\left(\mathcal{H}_{\bar A_i}\right)$. By considering the Stinespring dilation $V: \mathcal{H}_1 \to \mathcal{H}_1 \otimes \mathcal{H}_1$ defined by
\begin{align}
V \left(\ket{\psi_i}^{A_i} \otimes \ket{\phi_i}^{\bar A_i}\right) =  \ket{\psi_i}_{A_i} \otimes \ket{\omega_i}_{\bar A_i A_i} \otimes\ket{\phi_i}_{\bar A_i}
\end{align}
for some maximally entangled state $\ket{\omega_i} \in \mathcal{H}_{\bar{A}_i} \otimes \mathcal{H}_{A_i}$, we see that $\mathcal{P}_{\mathcal{A}'}$ is the complementary channel to $\mathcal{P}_{\mathcal{A}}$.

This relationship allows the application of the following theorem \cite{beny2010general}, defined in terms of the fidelity between channels
\begin{align}
F(\mathcal{N},\mathcal{M}) = \min_\rho F\left(\left(\mathcal{N} \otimes \Id\right)\rho,\left(\mathcal{M} \otimes \Id\right)\rho\right)
\end{align}
where $F(\rho,\sigma)$ is the usual fidelity between states.
\begin{thrm}\label{thrm:fid}
\begin{align}
\max_{\mathcal{R}} F \left( \mathcal{R} \circ \mathcal{N}, \mathcal{M}\right) = \max_{\mathcal{R}'} F \left( \mathcal N^c, \mathcal{R}' \circ \mathcal{M}^c\right)
\end{align}
where $\mathcal{Q}^c$ is the complementary channel to $\mathcal{Q}$.
\end{thrm}
If we take the channel $\mathcal{M}$ to be the projector $\mathcal{P}_{\mathcal A}$ onto the algebra and convert the statements about fidelities into statements about distances with respect to the diamond norm, we obtain the following result, which was first proved in \cite{beny2009conditions}.
\begin{thrm} \label{thrm:alinfdist}
Let 
\begin{align} \label{eq:errorcorrectalgebra}
\varepsilon_{\mathcal A} = \min_{\mathcal D} \left\lVert \mathcal{D} \circ \mathcal{N} - \mathcal{P}_{\mathcal{A}} \right \rVert_\diamond
\end{align}
and let
\begin{align} \label{eq:uncertaintyforgetalgebra}
\delta_{\mathcal A} = \left\lVert \mathcal N^c - \mathcal N^c \circ \mathcal P_{\mathcal A'} \right \rVert_\diamond. 
\end{align}
Then
\begin{align}
\frac{1}{4} \delta_{\mathcal A}^2 \leq \varepsilon_{\mathcal A} \leq 2 \delta_{\mathcal A}^{\frac{1}{2}}.
\end{align}
\end{thrm}
For the purposes of this paper, we are mostly content to consider the case of a single geometry, which could treat as subsystem quantum error correction without causing serious problems.\footnote{In reality, all the statements in Section \ref{sec:entwedge} are more precisely interpreted as statements about operator algebras; the use of the subsystem error correction paradigm was purely for pedagogical reasons. Since nothing in Section \ref{sec:entwedge} relied on the (false) assumption that the algebras had trivial centres, the entirety of Section \ref{sec:entwedge} can, if desired, be converted back into the (less sloppy) language of operator algebras using the dictionary given below.} This corresponds to the special case of operator algebra error correction where the centre of the algebra is trivial and hence the Hilbert space $$\mathcal{H}_1 = \mathcal{H}_a \otimes \mathcal{H}_{\bar a}.$$ The algebra $\mathcal A$ is simply operators on $\mathcal{H}_a$ while the commutant $\mathcal A'$ consists of operators on $\mathcal{H}_{\bar{a}}$. The projectors become
$$\mathcal P_{\mathcal A} = \Tr_{\bar a} \left(\cdot\right) \otimes \omega^{\bar a} \,\,\,\, \text{and} \,\,\,\, \mathcal P_{\mathcal A} = \omega^{a} \otimes 
 \Tr_{a} \left(\cdot\right),$$
 where $\omega$ is maximally mixed in both cases. It follows that
 $$ \varepsilon_{\mathcal A} = \min_{\mathcal D} \left\lVert \mathcal{D} \circ \mathcal{N} - \Tr_{\bar a} \left(\cdot\right) \otimes \omega^{\bar a} \right \rVert_\diamond = \min_{\mathcal D} \left\lVert \mathcal{D} \circ \mathcal{N} - \Tr_{\bar a} \left(\cdot\right)\right \rVert_\diamond.$$ 
 
If, as in Section \ref{sec:entwedge}, we define $\mathcal{N}$ to be the restriction of $\Tr_{\bar{A}}(\cdot)$ to $S(\mathcal{H}_{\text{code}})$ (and hence $\mathcal{N}^c$ is the restriction of $\Tr_A (\cdot)$ to $S(\mathcal{H}_{\text{code}})$), then $\varepsilon_{\mathcal A}$ and $\delta_{\mathcal A}$ become $\delta_1$ and $\delta_2$, as defined in (\ref{eq:epsilon}) and (\ref{eq:delta}), respectively. We see that Theorem \ref{thrm:alinfdist} becomes \eqref{eq:deltaepsilon}. 
 
As we saw throughout this paper, for large code spaces the fact $\delta_{\mathcal{A}}$ is defined using the diamond norm (and hence that we have to consider states entangled with a reference system) is crucially important. However, if the dimension $d$ of the code space is fixed in the limit $G_N \to 0$, we can use the bound \cite{hayden2012weak}
\begin{align}
\left\lVert \mathcal{N} - \mathcal{M} \right\rVert_\diamond \leq d \sup_{\ket{\psi} \in \mathcal{H}_{\text{code}}} \left\lVert \mathcal{N}(\psi) - \mathcal{M}(\psi) \right\rVert_1
\end{align}
to avoid considering any sort of reference system at all.

It was shown in \cite{dong2018entropy} that for any two, up to non-perturbative corrections, the boundary relative entropy,
\begin{align} \label{eq:aitor}
S(\rho_{\bar A} || \sigma_{\bar A})  = \langle A^{X_\sigma} + K_{\text{bulk},\sigma} \rangle_\rho - \langle A^{X_\rho} + K_{\text{bulk},\rho} \rangle_\rho,
\end{align}
where $A^{X}$ is the area of the RT surface, $K_{\text{bulk}}$ is the bulk relative entropy for region $\bar a$ and we have ignored higher curvature corrections (we assume $\lambda \to \infty$). Since the right hand side of \eqref{eq:aitor} depends only on the restriction of the bulk states to algebra $\mathcal{A}'$,
\begin{align}
S(\rho_{\bar A} || (\rho_{\mathcal{A}'})|_{\bar A} ) \leq \varepsilon,
\end{align}
for some $\varepsilon$ that is non-perturbatively small in $G_N$.\footnote{To be as rigorous as possible in our construction we should acknowledge that the boundary theory also does not actually factorise into Hilbert spaces on $A$ and $\bar A$ and hence $\mathcal{N}$ and $\mathcal{N}^c$ are really projectors onto the algebras of regions $A$ and $\bar A$ respectively, rather than partial traces onto subsystems. However, this makes no difference to the argument.} As usual, we convert the relative entropy into a trace distance using Pinsker's inequality, which gives
\begin{align} \label{eq:forgotten}
\left\lVert \left(\mathcal{N}^c - \mathcal{N}^c \circ \mathcal{P}_{\mathcal{A}'}\right) \rho \right\rVert_1 \leq \sqrt{2 \varepsilon \ln 2},
\end{align}
which is also non-perturbatively small. If the dimension $d$ of the code space is held fixed (or even grows polynomially in $1/G_N$ or $N$), then $\delta_{\mathcal{A}}$ and $\varepsilon_{\mathcal{A}}$ will also be non-perturbatively small and the error correction will be exact to all orders in perturbation theory

As a final technical comment, when working perturbatively to higher orders in $G_N$, we need to take into account the state dependence of the entanglement wedge $\bar{a}$. At finite $G_N$, different states therefore have different algebras $\mathcal{A}$ and $\mathcal{A}'$ associated to the same boundary regions $A$ and $\bar A$. To fix this we should instead work with the bulk algebra $\mathcal{A}'_0$  associated to the \emph{union} of the entanglement wedge $\bar{a}$ for all states in the code space. Then for any algebra $\mathcal{A}'$ associated to the entanglement wedge $\bar{a}$ of a particular state $\rho$,
\begin{align}
\mathcal{P}_{\mathcal{A}'}\mathcal{P}_{\mathcal{A}'_0}= \mathcal{P}_{\mathcal{A}'},
\end{align}
and hence
\begin{align}
\left\lVert \mathcal{N}^c \circ (\Id - \mathcal{P}_{\mathcal{A}'_0} ) \,\rho \right \rVert_1 \leq \left\lVert \mathcal{N}^c \circ (\Id - \mathcal{P}_{\mathcal{A}'} ) \,\rho \right \rVert_1 + \left\lVert \mathcal{N}^c \circ (\Id - \mathcal{P}_{\mathcal{A}'} ) (\mathcal{P}_{\mathcal{A}'_0} (\rho)) \right \rVert_1 \leq \sqrt{8 \varepsilon \ln 2}
\end{align}
where we have used the fact that both $\rho$ and $\mathcal{P}_{\mathcal{A}'_0} (\rho)$ have the same entanglement wedge and hence satisfy \eqref{eq:forgotten} for the same algebra $\mathcal{A}'$. 

It follows that only the algebra $\mathcal{A}_0$ associated to the \emph{intersection} of the entanglement wedge $a$ for all states can be decoded from region $A$. In the limit $G_N \to 0$, the entanglement wedge for every state converges to the classical entanglement wedge with no backreaction and hence the algebra $\mathcal{A}_0$ becomes the algebra associated to the classical region $a$.
 
 \bibliographystyle{JHEP}
\bibliography{biblio}
\end{document}